\documentclass[sigplan,10pt]{acmart}
\renewcommand\footnotetextcopyrightpermission[1]{}
\usepackage{textcomp}
\usepackage{algorithm}
\usepackage[normalem]{ulem}
\usepackage{algpseudocode}
\usepackage{subfig}
\usepackage{array}
\usepackage{ulem}

\setcopyright{rightsretained}

\acmDOI{10.475/123_4}

\acmISBN{123-4567-24-567/08/06}

\acmConference[Middleware'21]{22nd ACM/IFIP International Conference}{December 2021}{Québec City, Québec, Canada}
\acmYear{2021}
\copyrightyear{2021}

\acmArticle{4}
\acmPrice{15.00}

\begin{document}

\title{Memory at Your Service: Fast Memory Allocation for Latency-critical Services}

\author{Aidi Pi, Junxian Zhao, Shaoqi Wang, Xiaobo Zhou}
\affiliation{%
  \institution{University of Colorado Colorado Springs}
  \streetaddress{}
  \postcode{}
}
\email{{epi, jzhao, swang, xzhou}@uccs.edu}

\begin{abstract}
Co-location and memory sharing between latency-critical services, such as key-value store and web search, and best-effort batch jobs is an appealing approach to improving memory utilization in multi-tenant datacenter systems. However, we find that the very diverse goals of job co-location and the GNU/Linux system stack can lead to severe performance degradation of latency-critical services under memory pressure in a multi-tenant system.

We address memory pressure for latency-critical services via fast memory allocation and proactive reclamation. We find that memory allocation latency dominates the overall query latency, especially under memory pressure. We analyze the default memory management mechanism provided by GNU/Linux system stack and identify the reasons why it is inefficient for latency-critical services in a multi-tenant system. We present Hermes, a fast memory allocation mechanism in user space that adaptively reserves memory for latency-critical services. It advises Linux OS to proactively reclaim memory of batch jobs. We implement Hermes in GNU C Library. Experimental result shows that Hermes reduces the average and the $99^{th}$ percentile memory allocation latency by up to 54.4\% and 62.4\% for a micro benchmark, respectively. For two real-world latency-critical services, Hermes reduces both the average and the $99^{th}$ percentile tail query latency by up to 40.3\%. Compared to the default Glibc, jemalloc and TCMalloc, Hermes reduces Service Level Objective violation by up to 84.3\% under memory pressure.
\end{abstract}

\settopmatter{printfolios=true}

\maketitle

\section{Introduction}
Latency-critical services such as key-value store and web search are usually featured with largely varied peak and average resource consumption~\cite{Stuedi-ATC19-Crail, Hao-SOSP17-MittOS}. For guaranteed performance of latency-critical services, a naive approach is to use a dedicated system for latency-critical services. However, the approach leads to a large amount of idle resources during runtime since the average resource consumption of the services is usually much less than their peak consumption~\cite{fried2020caladan, Leverich-EuroSys14-Reconciling, Vuppalapati-NSDI20-SnowFlake}. For instance, SnowFlake system found that the average memory utilization on its servers is only $\sim19\%$~\cite{Vuppalapati-NSDI20-SnowFlake}. In order to improve the utilization of resources, it is a common practice that best-effort batch jobs are co-located with latency-critical services to exploit transient resources in datacenters~\cite{fried2020caladan, Iorgulescu-ATC18-PerfIso, Yang-ATC16-Elfen, Zhu-ASPLOS16-Dirigent, Javadi-ASPLOS16-DIAL, Yang-ISCA13-Bubble-Flux}.

Although co-location with memory sharing increases resource utilization, it often significantly degrades the latency particularly the tail latency of latency-critical services. 
Latency-critical services like web search commonly distribute requests across many servers, thus the end-to-end response time is determined by the slowest individual latency~\cite{TailAtScale,Zhu-SOCC17-WorkloadCompactor,Berger-OSDI18-RobinHood,fried2020caladan}.
The root cause of long tail latency is due to the very diverse goals of job co-location and the GNU/Linux system stack. On one hand, job co-location leverages idle resources for batch jobs while maintaining the performance of latency-critical services. On the other hand, the GNU/Linux stack tries to accommodate as many submitted processes as possible while only offering few knobs to prioritize processes. As a result, although co-located latency-critical services and batch jobs may both survive, the performance of latency-critical services is significantly degraded under memory pressure, which jeopardizes Service Level Objective (SLO).

There are mainly two categories of research on improving performance for latency-critical services. Studies~\cite{Berger-OSDI18-RobinHood, Li-PPoPP16-tailcontrol, Zhu-SOCC17-WorkloadCompactor, ROLP-EuroSys19-Bruno} improve performance for latency-critical services by leveraging their runtime characteristics. For example, ROLP~\cite{ROLP-EuroSys19-Bruno} is a runtime object lifetime profiler for efficient memory allocation and garbage collection for latency-critical services. However, these studies do not take job co-location into consideration. Studies of the other category~\cite{Iorgulescu-ATC18-PerfIso, Zhu-ASPLOS16-Dirigent, Javadi-ASPLOS16-DIAL, Yang-ISCA13-Bubble-Flux} target co-location of latency-critical services with other jobs. For example, PerfIso~\cite{Iorgulescu-ATC18-PerfIso} and Dirigent~\cite{Zhu-ASPLOS16-Dirigent} are two representative approaches that leverage multicore systems to efficiently share CPU resource between processes. 
Our work falls into the second category.

While existing efforts try to push the resource utilization to the limit, memory management for latency-critical services still faces significant challenges. First, the runtime behavior of a job is difficult to predict. In particular, it is difficult to obtain the amount of memory that will be requested by a job in the future. Second, it is expensive to reclaim physical memory that is occupied by a process. If a process requests more memory when the node memory is almost used up, swapping will be triggered to make space for the requested memory. However, swapping is an expensive operation that takes a long period of time (tens of milliseconds to seconds) or even leads to thrashing. In such cases, the performance of latency-critical services are significantly degraded.

Since the original purpose of a dedicated system is for sole use by latency-critical services, ideally their performance should not be affected by batch jobs. In a shared environment, memory is frequently allocated and reclaimed due to provisioning of various  workloads. However, the memory reclaim mechanism in Linux OS significantly degrades the performance of latency-critical services under memory pressure, which makes co-location inefficient or even ineffective. In light of the challenges, we tackle the problem from a new perspective:  
resource slacks should be reserved for latency-critical services in case of a burst of resource requests.  
PerfIso~\cite{Iorgulescu-ATC18-PerfIso} is a preemptive approach that adopts this principle to achieve CPU sharing between latency-critical services and batch jobs. 
However, data in memory can only be preempted by swapping them onto disks, which is a very expensive operation.
We aim to materialize the principle to achieve fast memory allocation for latency-critical services in a multi-tenant system. Our experiments find that memory allocation latency takes up to 97.5\% of a whole query latency. Thus, we focus on reducing the memory allocation latency for latency-critical services. The design should meet the following requirements: 
\begin{itemize}
\item \textbf{R1} Latency-critical services have the highest priority. This is the primary principle. Best-effort batch jobs can share idle resources only if they do not affect the performance of latency-critical services.

\item \textbf{R2} Memory should be allocated in a fast manner. This is the key to achieving low latency for latency-critical services when they request memory. 

\item \textbf{R3} The design should be generally applicable to all applications written in a popular language such as C / C++. That is, the source code of applications should not be modified.

\item \textbf{R4} The overhead should be low. In other words, it should consume little resource of a node.
\end{itemize}

In this paper, we make the following contributions. First, we analyze the current memory management in GNU C Library (a.k.a. Glibc) and Linux OS, and show that it is inefficient for memory sharing between latency-critical services and batch jobs. In particular, 1) it adopts an on-demand physical memory allocation mechanism in order to accommodate as many processes as possible without prioritization. Though this mechanism works well in a dedicated system with sufficient memory, it significantly degrades job performance or even causes thrashing under memory pressure. 2) It uses a reactive algorithm to reclaim file cache even if no process accesses the cache. The design expects the cache will be accessed again in the near future. The reactive algorithm introduces significant delay on latency-critical services since a memory reclaim routine is invoked before requests are served. In summary, the design of the current GNU / Linux stack contradicts the goal of co-location and memory sharing of latency-critical services and batch jobs.

Second, we present Hermes, a library-level mechanism for fast memory allocation for latency-critical services in multi-tenant systems. Hermes maintains one dedicated memory pool for each latency-critical service (\textbf{R1}, \textbf{R2}). Upon receiving requests from a latency-critical service, memory can be immediately allocated from the memory pool to the service. Hermes uses a lightweight heuristic to determine the size of the memory pool (\textbf{R4}). It advises Linux OS to release file cache pages occupied by batch jobs under memory pressure so as to make more available memory for latency-critical services (\textbf{R1}). We implement Hermes in library Glibc. It is a library-level mechanism without modification to applications (\textbf{R3}) or Linux OS. Note that Hermes could be implemented into Linux OS, but the modification may affect other processes, incur security issues, and importantly violate Linux monolithic kernel generality.

We conduct experiments for Hermes with a micro benchmark and two real-world services under a multi-tenant system. 
Compared to the default Glibc, Hermes reduces the average and the $99^{th}$ percentile memory allocation latency by up to 54.4\% and 62.4\% under memory pressure, respectively.
The allocation latency is as low as $4 \mu s$ for small requests and $1 ms$ for large requests. We use Redis~\cite{Redis} and Rocksdb~\cite{Rocksdb} as two real-world services to examine the query latency. Results show that Hermes reduces both the average and the $99^{th}$ percentile tail query latency by up to 40.3\%. Compared to the default Glibc, jemalloc and TCMalloc, Hermes reduces the SLO violation by up to 84.3\% under memory pressure. Hermes achieves significantly improved system throughput. Results also show that Hermes achieves similar or slightly better query latency under a dedicated system. The overhead of Hermes is negligible.


\section{Background and Motivations}
\label{sec:motivation}

\subsection{Memory Management in Glibc}
\label{sec:glibc-bg}
\begin{figure}
\centering
\includegraphics[width=\linewidth]{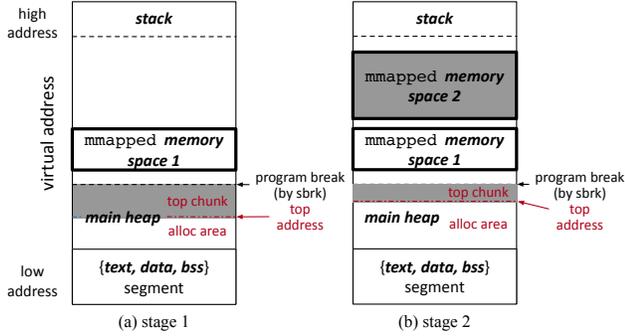}
\caption{Process address space in Linux. Shaded areas represent allocated virtual memory whose physical pages do not reside in RAM. Red fonts represent variables on Glibc.}
\label{fig:address-space}
\end{figure}

The famous \texttt{malloc} function call in Glibc is a unified interface for programs to allocate memory from Linux OS. A process conveniently obtains the address of the memory space without knowing the underlying mechanism by calling \texttt{malloc}. The function call uses two Linux system calls \texttt{brk} and \texttt{mmap} to serve memory requests of different sizes. Figure~\ref{fig:address-space}(a) shows the simplified address space of a process that includes memory chunks allocated by both system calls. We focus on the mechanisms in Glibc that manipulate the main heap space and mmapped memory chunks. Both kinds of memory are dynamically allocated at runtime.

\textbf{System call \texttt{brk}.} Each process has exactly one main heap that is a continuous virtual address space. Glibc divides the main heap into two areas: the allocated area and the top chunk. Glibc keeps track of the used and free space in the allocated area. It is worth noting that the allocated area and the top chunk in Glibc are transparent to Linux OS. Following the allocated area lies the top chunk that is a continuous free address space. The end address of the top chunk is the program break returned by the \texttt{sbrk} wrapper function which calls the \texttt{brk} system call. Upon a request for a small size of memory ($<$ 128 KB by default), Glibc first tries to find a free space in the allocated area. If it cannot satisfy the request, space is taken from the beginning of the top chunk and added to the allocated area. Once the top chunk is used up, Glibc expands the main heap by calling \texttt{sbrk} with the exact requested size. If the top chunk is greater than a certain threshold, Glibc shrinks the main heap by passing a negative number to \texttt{sbrk}. 

\textbf{System call \texttt{mmap}.} Besides the main heap, a process can have multiple disjoint memory chunks allocated by \texttt{mmap}. This system call can either map a file to process address space or allocate anonymous pages. Glibc leverages the anonymous page usage to handle large memory requests ($\geq$ 128 KB by default). Upon success, it returns the starting address of the newly allocated mmapped memory chunk. Glibc gives the memory chunk to the process after a bookkeeping operation. When a process frees a memory space allocated by \texttt{mmap}, Glibc releases it directly back to Linux OS.

Upon return of both system calls, a process gets a virtual memory space while the corresponding physical memory does not necessarily reside in RAM at the moment. Linux OS constructs the virtual-physical address mapping only when the process accesses (i.e., writes or reads or executes) the allocated memory for the first time. For example, in Figure~\ref{fig:address-space}(a), the process has a main heap and a mmapped memory space 1. In Figure~\ref{fig:address-space}(b), the process allocates a new mmapped memory space and writes data in the main heap. The newly mmapped memory space does not have corresponding physical pages yet, and the virtual-physical mapped space in the main heap expands. Two benefits come with the on-demand mapping construction. For Linux OS, physical memory pages are loaded for the actually used memory since physical memory is a scarce resource. For the process, it accelerates the memory allocation routine. The reason is that the mapping construction for all the virtual addresses requires loading all the physical pages at once, which takes a longer time than only returning the virtual address.

While usually fast, the on-demand virtual-physical mapping construction can be significantly delayed when there is insufficient physical memory in the node, which is common in a multi-tenant system. At this point, Linux OS starts to reclaim physical pages by either directly freeing them or swapping them onto disks.


\begin{figure}
\centering
\subfloat[Small (1KB) requests.]{
\includegraphics[width=0.24\textwidth]{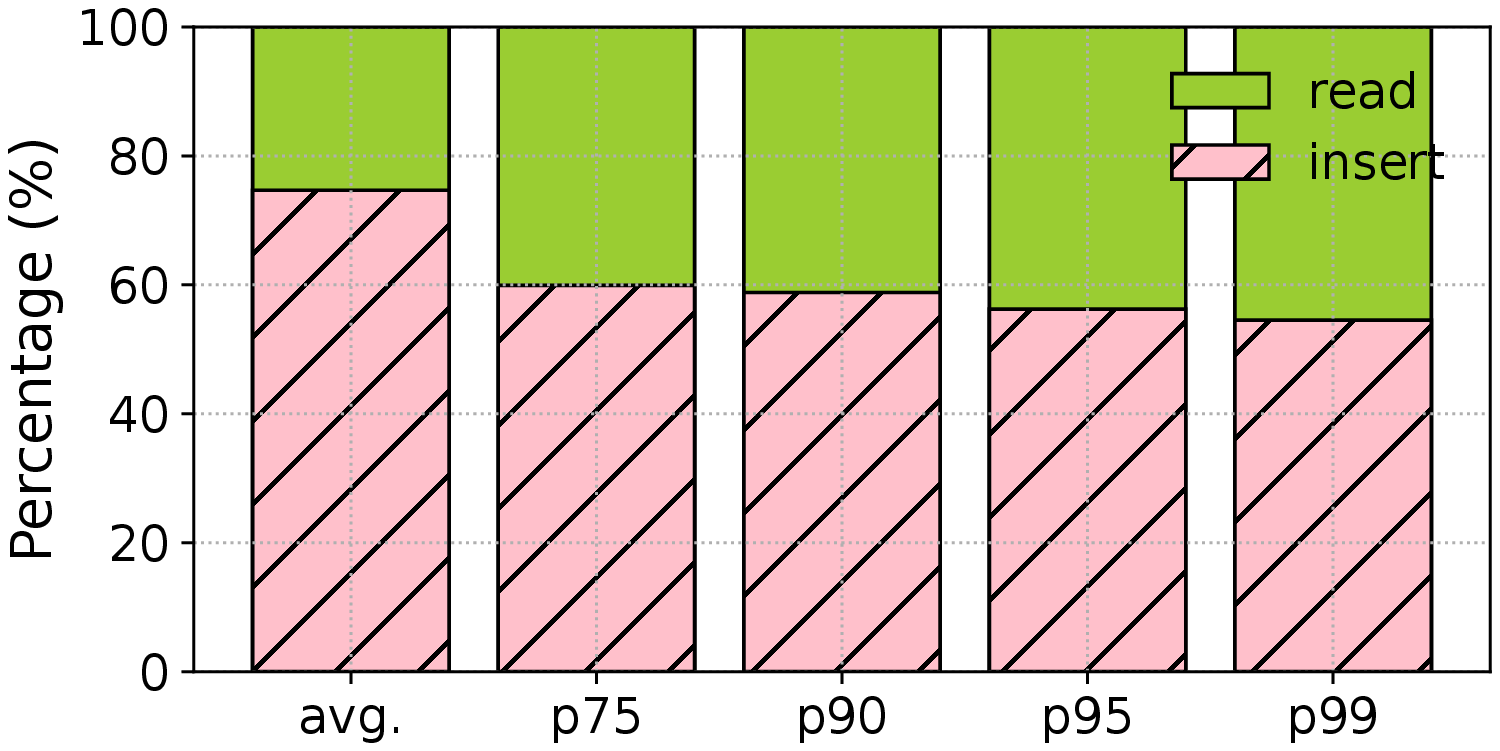}
}
\subfloat[Large (200KB) requests.]{
\includegraphics[width=0.24\textwidth]{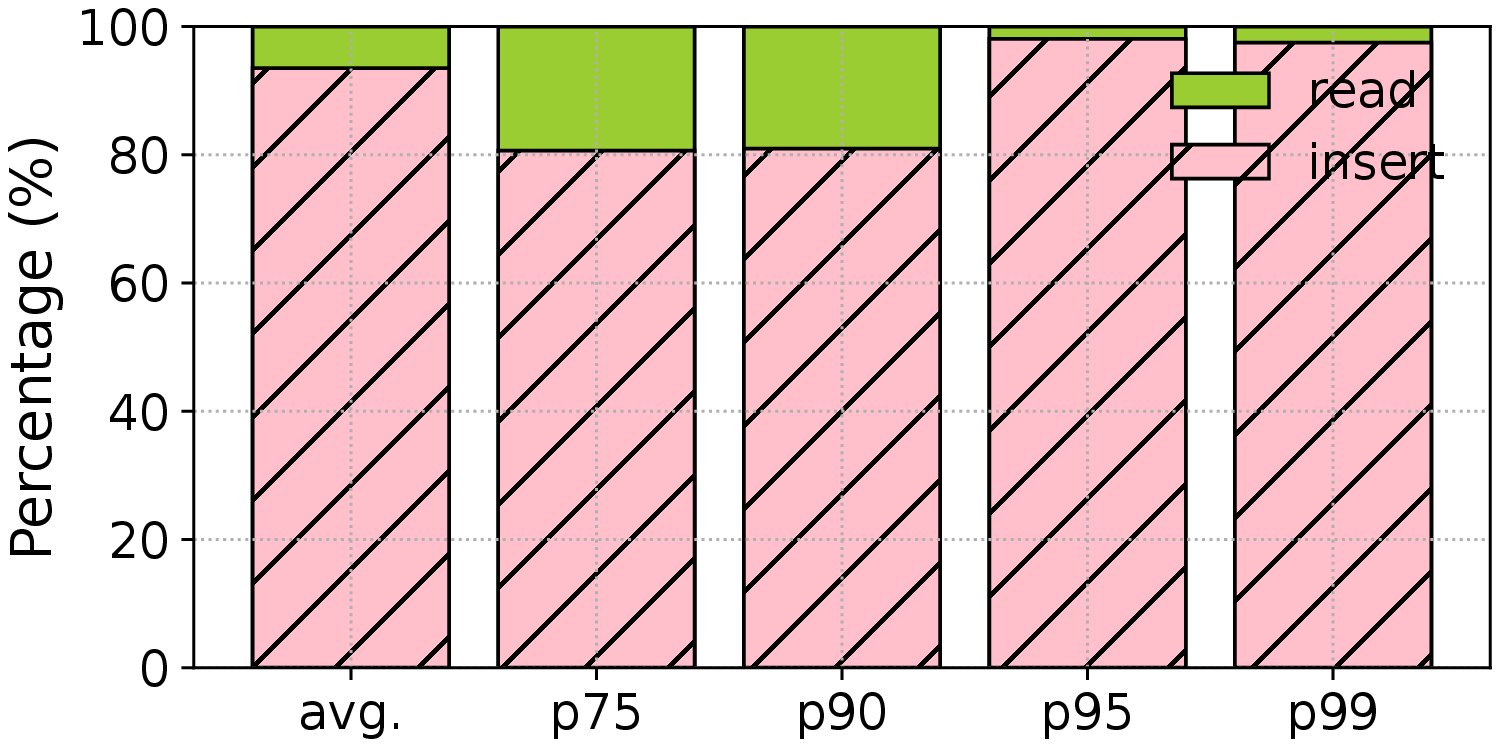}
}
\caption{Percentage breakdown of the insert and read operations in Rocksdb.}
\label{fig:allocation-ratio}
\end{figure}

\subsection{Case Studies}
In real-world latency sensitive services, latency spent in memory allocation during data insertion takes a large portion of latency of a whole workload. We take Rocksdb as a case study to illustrate that memory allocation latency is much higher compared to data read latency using both small (1KB) and large (200KB) requests. We use Glibc as the memory allocator andexecute Rocksdb queries without any memory pressure. Each query is a data insertion operation (involving memory allocation) followed by a read operation.  Figure~\ref{fig:allocation-ratio} shows the percentage breakdown of the query latency at specified percentiles. For small requests, the  data insertion latency is 74.7\% (54.5\%) of the average ($99^{th}$ percentile) overall query latency. For large request, the ratio is 93.5\% (97.5\%) of the average ($99^{th}$ percentile) overall query latency. The impact of memory allocation is significant, and even more in large requests. As for data update requests, it renders similar results compared with read quests since they do not incur memory allocation.

\begin{figure}
\centering
\includegraphics[width=0.8\linewidth]{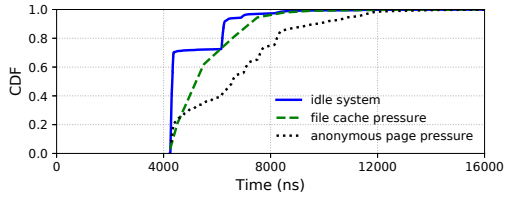}
\caption{CDF of the memory allocation latency. }
\label{fig:motivation}
\end{figure}

We use another case study to demonstrate the memory allocation latency degradation under anonymous page pressure and file cache pressure. We use a micro benchmark that continuously sends 1KB-size memory requests until a total amount of 1 GB, using the default Glibc in a node with 128 GB RAM. We repeat the experiment under a dedicated system with sufficient memory, under anonymous page pressure, and under file cache pressure, respectively. The details of the micro benchmark and the node are described in Section~\ref{sec:evaluation_setup}. Figure~\ref{fig:motivation} shows the CDF of the memory allocation latency under the dedicated system and two kinds of memory pressure.

\noindent
\textbf{Anonymous page pressure.} To generate anonymous page pressure, we run a program that continuously sends memory allocation requests until the available memory in the node becomes about 300 MB. Note that, the available memory could not further drop below 300 MB due to the indirect and direct reclaim mechanisms of Linux OS. At this point, new memory allocation requests from the micro benchmark trigger the memory reclaim routine and cause swapping. Figure~\ref{fig:motivation} shows that the memory allocation latency significantly increases under anonymous page pressure. The average and the $99^{th}$ percentile allocation latency under anonymous page pressure are prolonged by 35.6\% and 46.6\% compared to those without memory pressure, respectively.

\noindent
\textbf{File cache pressure.} We generate file cache pressure by loading 10 GB files and sending memory allocation requests to occupy the rest of the system memory until free memory drops to about 300 MB. In this case, memory reclaim routine starts but not necessarily trigger swapping since the file cache can be directly released without accessing the disk. Figure~\ref{fig:motivation} shows that the memory allocation latency under file cache pressure is lower than that under anonymous page pressure, but it is still higher than that under a dedicated system. The average and the $99^{th}$ percentile allocation latency under file cache pressure are prolonged by 10.8\% and 7.6\% compared to those without memory pressure, respectively.

Memory pressure significantly prolongs memory allocation latency, which has non-trivial impact on SLO violation. We target on both kinds of memory pressure and aim to reduce the memory allocation latency of latency-critical services in a co-located system as well as in a dedicated system.

\subsection{Memory Reclaim in Linux OS}
Linux OS emulates an LRU-like (Least Recent Used) algorithm for physical memory page reclaim by keeping four lists: \texttt{active\_anon} and \texttt{inactive\_anon} for anonymous pages, and \texttt{active\_file} and \texttt{inactive\_file} for file cache pages. The two active lists contain recently used pages while the two inactive lists contain pages that are not recently used. Under memory pressure, Linux OS scans through these four lists, updates page usage status, moves pages between lists, and selects pages to reclaim. Specifically, Linux OS keeps three memory wartermarks (i.e. high, low and minimum) to instruct memory reclaim routine. When available memory drops below the low watermark, a page reclaim thread is started until available memory is larger than the high watermark. When available memory further drops below the minimum watermark, each memory request goes through a synchronous direct memory reclaim routine before the physical memory is allocated. 

\begin{figure}
\centering
\includegraphics[width=0.82\linewidth]{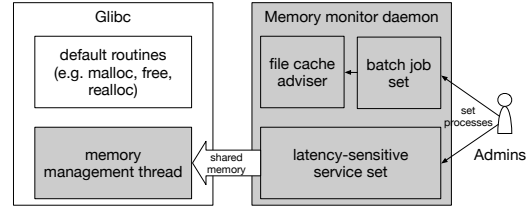}
\caption{The architecture of Hermes.}
\label{fig:overview}
\end{figure}

However, the page reclaim algorithm in Linux OS is inefficient for latency-critical services in a multi-tenant system. The watermarks are conservatively set at around 1{\textperthousand} of a memory zone. For example, the total capacity of a memory zone in one of our physical nodes is 60 GB. The low and high watermarks are 53 MB and 64 MB, respectively. Since both latency-critical services and batch jobs tend to consume hundreds of megabytes or gigabytes of memory, the watermarks are too small to timely trigger the indirect memory reclaim thread. The direct memory reclaim routine even causes more delays on memory requests. After a process finishes, all of its anonymous pages are reclaimed immediately. However, the file cache pages loaded by the process are not reclaimed by Linux OS but remain in memory. They are only reclaimed upon memory pressure by the reclaim routine, which prolongs new memory requests. The memory pressure cannot be relived even if we increase the watermarks. Although, Linux OS triggers memory reclaim routine when there is still much free memory with higher watermarks, it does not distinguish latency-critical services and batch jobs. Memory from both kinds of workloads can be reclaimed. The performance of latency-critical services is still degraded.


\noindent
\textbf{[Summary]} There are two drawbacks of the current GNU / Linux system stack that make the memory allocation of latency-critical services inefficient in a multi-tenant system. 1) Glibc only keeps a small chunk of physically mapped memory in the main heap, which is much less than the total size of memory requests from latency-critical services. 2) The on-demand virtual-physical memory mapping construction causes significant delay under memory pressure due to the conservative memory page reclaim mechanism in Linux OS. 

\begin{figure}
\includegraphics[width=0.82\linewidth]{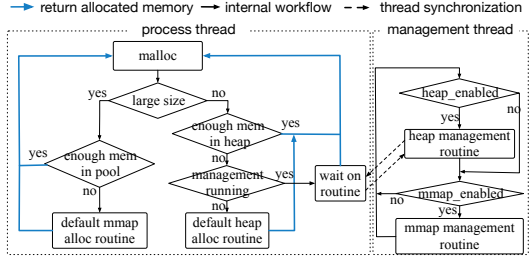}
\caption{The workflow of the modified Glibc routines.}
\label{fig:management_thread}
\end{figure}

\section{Hermes Design}
\label{sec:design}
\subsection{Overview}

%

In this paper, we propose and develop Hermes, a library-level mechanism to memory management that addresses the identified problems in the GNU/Linux system stack and reduces memory allocation latency of latency-critical services in a multi-tenant system. Hermes is transparent to applications and it does not make modification to Linux OS. As shown in Figure~\ref{fig:overview}, Hermes consists of two major components: a memory management thread woken per $f$ milliseconds in Glibc and a memory monitor daemon independently running on the same physical node. A system administrator sends the process IDs of batch jobs and latency-critical services to the memory monitor daemon. Upon memory pressure, the file cache adviser advises Linux OS to free the file cache owned by batch jobs. In Glibc, if a process is a latency-critical service, the memory management thread is started for memory reservation and virtual-physical address mapping.

\subsection{Memory Management Thread}
The goal of the memory management thread is to reserve memory and construct its virtual-physical address mapping in advance for latency-critical services. Figure~\ref{fig:management_thread} outlines the workflow of the management thread and the modified Glibc. The management thread periodically checks the current amount of reserved memory and decides whether to reserve more memory or release reserved memory back to Linux OS. When a process thread calls \texttt{malloc}, Hermes first tries to return the reserved memory to the process. If the reserved memory is insufficient, it uses the default routine to serve the request. Though sharing the same principle, the management thread uses different approaches to manage the main heap memory and mmapped memory chunks since they are allocated by two different system calls.
%

\subsubsection{Heap Memory Management}
Small-sized memory requests are allocated from the main heap, as shown in the \texttt{no} branch of the \texttt{large\_size} statement in Figure~\ref{fig:management_thread}. If there is sufficient memory in the main heap, Hermes immediately allocates it to the requests. Otherwise, if the management thread is running, the requests wait on it. If memory in the main heap is insufficient, the requests are allocated by the default allocation routine in Glibc. We show the heap management routine in Algorithm~\ref{alg:heap}. In every round of the execution, the routine first updates the memory allocation metrics including the total size of all small memory requests (i.e. requests $<$128 KB) and the number of requests in the last interval. It then updates all the thresholds based on the collected memory allocation metrics (function \textsc{UpdateThreshold}). For example, the target amount of reserving memory is the total amount of memory requests in the last interval multiplying a reservation factor $RSV\_FACTOR$. If the top chunk is smaller than the reservation threshold $RSV\_THR$, it expands the current program break and immediately constructs the virtual-physical mapping for the newly allocated memory. Otherwise, if the free space in the top chunk exceeds the trim threshold $TRIM\_THR$, it shrinks the top chunk by setting the program break to a lower memory address.


\begin{figure}
\centering
\includegraphics[width=\linewidth]{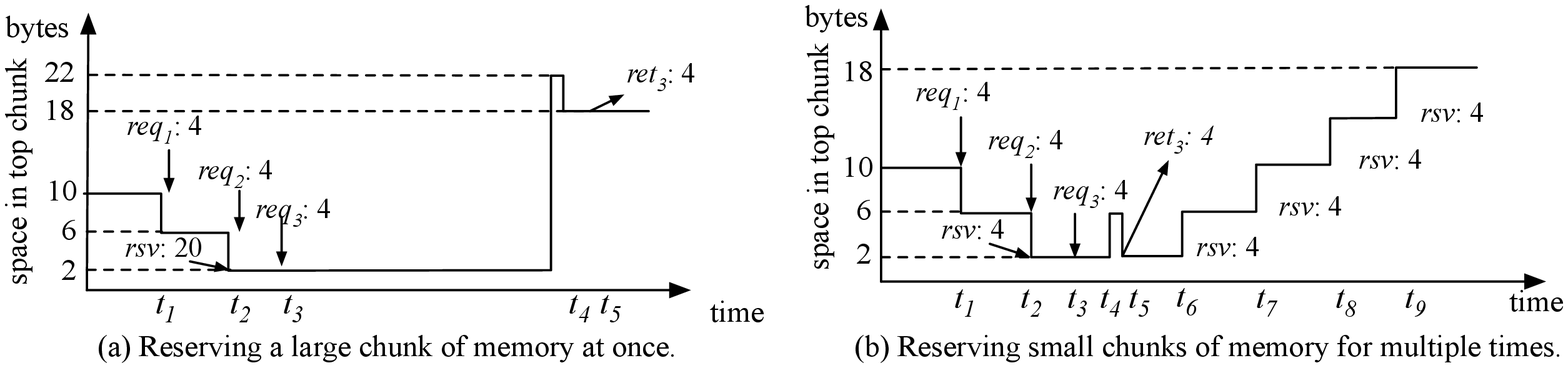}
\caption{Illustration of gradual reservation.}
\label{fig:small_chunk}
\end{figure}

\begin{algorithm}[H]
\scriptsize
\caption{Heap management routine.}
\label{alg:heap}
\begin{algorithmic}[1]
\State $RSV\_THR$: a threshold below which more memory should be reserved;
\State $TGT\_MEM$: the target free size in the top chunk at which the memory reservation stops;
\State $TRIM\_THR$: a threshold above which memory is released;
\State $MEM\_CHUNK$: memory reserved on each \texttt{sbrk()} call;
\State $top\_free$: current free memory in the top chunk;
\State \Call{UpdateThreshold}{ };
\If{$top\_free < RSV\_THR$}
	\State $mem\_to\_reserve \gets (TGT\_MEM - top\_free)$;
	\State $reserved \gets 0$;
	\While{$reserved < mem\_to\_reserve$}
		\State \Call{Lock}{heap};
		\State $address \gets$ \texttt{sbrk($MEM\_CHUNK$)};
		\State \Call{ConstructMapping}{$address$};
		\State $reserved \gets (reserved + MEM\_CHUNK)$;
		\State \Call{Unlock}{heap};
	\EndWhile
\ElsIf{$top\_free > TRIM\_THR$}
	\State $extra \gets (top\_free - TRIM\_THR)$;
	\State \Call{Lock}{heap};
	\State \texttt{sbrk($-extra$)};
	\State \Call{Unlock}{heap};
\EndIf
\end{algorithmic}
\end{algorithm}

\noindent
\textbf{A naive approach.} The challenge of expanding the main heap lies in how to determine the amount of memory to be reserved. Intuitively, simply reserving a large amount of memory at once would boost process performance since the memory is immediately available for processes. However, our experiments find that this approach even degrades the performance of latency-critical services in terms of tail latency. The latency of the default on-demand virtual-physical mapping construction is near proportional to the size of the constructed memory. Since there is only one program break for each process, the manipulation on the program break must be synchronized. 

A burst of memory requests in the process thread may be blocked for a long time due to the mapping construction for a large chunk of memory in the management thread. Figure~\ref{fig:small_chunk} (a) illustrates this scenario. There are initially 10 bytes in the top chunk. At $t_1$ and $t_2$, the user process sends two memory requests $req_1$ and $req_2$ of 4 bytes, respectively. The requests return immediately. Then, there are only 2 bytes left in the top chunk. The management thread is now invoked to expand the top chunk by 20 bytes and construct the virtual-physical mapping. At $t_3$, there is another request $req_3$ of 4 bytes from the user process. Since the running management thread locks the program break, $req_3$ is blocked. It can only be served at $t_5$ after the top chunk is expanded at $t_4$, which incurs significant delay on the request. Although a large number of memory requests do not compete with the main heap expansion, it is the competing ones that  lead to prolonged tail latency.

\noindent
\textbf{Gradual reservation.} We propose gradual reservation that expands the program break by a small size at a time for multiple times (lines 10 $\sim$ 16 in Algorithm~\ref{alg:heap}). For example, instead of expanding the program break for 20 bytes at once, gradual reservation expands the program break for 5 times, each time for 4 bytes, as shown in Figure~\ref{fig:small_chunk}(b). Before $req_3$ arrives, a reservation of a small memory chunk has already been sent to Linux OS at $t_2$ by the management thread. After the reservation returns, $req_3$ can be immediately served. Finally, the management thread sends four more small reservation operations until the reserved memory reaches 18 bytes. Based on our observation and other studies~\cite{Hao-SOSP17-MittOS, Bonwick-1994-Slab}, continuous memory requests from latency-critical services are usually of a similar or constant size. Hermes uses the average memory request size during the previous interval as the size of each memory chunk in gradual reservation. Compared with the default on-demand virtual-physical mapping construction, Hermes serves memory requests faster even if the program break is locked by the management thread, because the virtual-physical mapping construction already starts in advance and returns shortly.

\subsubsection{Mmapped Memory Management}
Large memory requests are allocated from mmapped memory chunks, as shown in the \texttt{yes} branch of the \texttt{large\_size} statement in Figure~\ref{fig:management_thread}. Management for mmapped memory is asynchronous since a process can have multiple chunks of mmapped memory space. In other words, the process thread and the management thread can simultaneously allocate two different chunks of mmapped memory space. Thus, incoming requests do not wait on the management thread but uses the default memory allocation routine when the reserved memory is insufficient. Algorithm~\ref{alg:mmap} shows the management routine for mmapped memory. Since the addresses of mmapped memory space are not necessarily adjacent, each chunk of space needs to be managed separately. We use a segregated free list as the memory pool to keep track of the addresses of mmapped memory space (line 14).
The function calculates the target bucket based on the size of a mmapped memory chunk using formula~\ref{eq: hash}. 

Parameter $min\_mmap\_size$ is the minimum memory request size that can use \texttt{mmap} system call, which is 128 KB by default in Glibc. Parameter $table\_size$ is the maximum number of buckets in the segregated free list. In implementation, we empirically set $table\_size$ to 8 (1 MB / 128 KB) since the size of a single memory request is usually less than 1 MB.
\begin{equation}
\label{eq: hash}
bucket(chunk\_size)= MIN (\Big\lfloor\frac{chunk\_size}{min\_mmap\_size}\Big\rfloor, table\_size)
\end{equation}


\begin{algorithm}[H]
\scriptsize
\caption{Mmap management routine.}
\label{alg:mmap}
\begin{algorithmic}[1]
\State $RSV\_THR$: a threshold below which more memory is reserved;
\State $TGT\_MEM$ the target free size of mmapped space at which reservation stops;
\State $TRIM\_THR$: a threshold above which memory is released;
\State $MEM\_CHUNK$: memory reserved on each \texttt{mmap()} call;
\State $memory\_pool$ a segregated free list that keeps track of the allocated mmapped space;
\State $alloc\_set$: a set of allocated mmapped chunks by the process thread;
\State \Call{DelayRelease}{$alloc\_set$};
\State \Call{UpdateThreshold}{ };
\If{$memory\_pool.total\_size < RSV\_THR$}
	\State $reserved \gets 0$;
	\While{$reserved < TGT\_MEM$}
		\State $address \gets$ \texttt{mmap($MEM\_CHUNK$)};
		\State \Call{ConstructMapping}{$address$};
		\State $memory\_pool.add(address)$;
		\State $reserved \gets (reserved + MEM\_CHUNK)$;
	\EndWhile
\EndIf
\While{$memory\_pool.total\_size > TRIM\_THR$}
	\State $to\_release \gets memory\_pool.smallest\_space$;
	\State \texttt{munmap($to\_release$)};
\EndWhile
\end{algorithmic}
\end{algorithm}

Upon a request for a large chunk of memory (i.e., requests $\geq$ 128 KB) from the process, the modified allocation routine first tries to find the best-fit bucket in the list by calculating the bucket based on the requested size. The hash code of the best-fit bucket is calculated by equation $MIN (bucket(request\_size) + 1, table\_size)$. If there is no such a chunk, the allocation routine uses the largest chunk in the memory pool and expands the chunk to the requested size. If this step still fails due to an empty memory pool, it falls back to the default allocation using \texttt{mmap} system call. For example, in Figure~\ref{fig:hash-table}, there are three memory chunks. Two are in bucket 1 and the third is in bucket 2. Consider that the application process sends a 278 KB-size memory request. The hash code of the best-fit bucket is 2. Hermes takes the first memory chunk (the 524 KB-size chunk) in the corresponding bucket and returns it to the application process. It is worth noting that, Hermes does not first choose bucket 1 as the target bucket since it may contain chunks that are smaller than the memory request. Otherwise, Hermes needs to scan through the memory chunks in the buckets in order to find a memory chunk that is larger than the requested size, which introduces more overhead. As a result, the allocated chunks are usually not exactly the same size as the request size. After returned to the user process, they are put into $alloc\_set$. On the next round execution of the management thread, they are shrunk to the size of the requests (line 7). By this design, the user process gets requested memory immediately as long as they are available while asynchronous shrinking avoids memory wastage. If memory requests are served by expanding an existing small chunk, the delay is still shorter than that of the default allocation routine. The reason is that small chunks already have their virtual-physical mapping constructed. Additional mapping constructions only need to be done for the space that exceeds the size of the original memory chunks.

\subsection{Memory Monitor Daemon}
The memory monitor daemon is running on a physical node that adopts job co-location.
The daemon keeps the process IDs of latency-critical services in shared memory. The memory management thread adopts a lazy initialization mechanism. When a process detects its process ID is in the shared memory, it initializes the memory management thread. Otherwise, the process behaves as it uses the default Glibc.

\noindent
\textbf{Proactive reclamation.} The memory monitor daemon is responsible for proactively advising Linux OS to release file cache pages upon memory pressure. The daemon keeps track of all batch jobs and their loaded data files. When the system memory usage exceeds threshold \texttt{adv\_thr}, the monitor daemon advises Linux OS to release file cache pages in a largest-file-first order until the percentage of file cache drops below the threshold or no file cache is from the specified batch jobs. 
The largest-file-first paging order makes a large chunk of memory available at once for latency-critical services. It also reduces the number of calls to the advising routine.

Proactive reclamation is an effective approach to accelerating memory allocation. Although Hermes reserves physical memory in advance, the reservation can still be delayed if it triggers the direct reclaim routine due to insufficient memory. Proactive reclamation reduces the chance by which the direct reclaim routine is triggered. Note that solely relying on proactive reclamation is insufficient since it only tries to make free space for new memory requests but it does not contribute to virtual-physical mapping construction.

\section{Implementation}
\label{sec:implementation}

We implement Hermes in Glibc-2.23 with about 1,200 lines of C code. We empirically set the invocation interval ($f$) of the memory management thread to 2. Recall that we use a reservation factor $RSV\_FACTOR$ to determine the amount of memory to be reserved. A larger value results in more reserved memory and faster memory allocation. However, the reserved memory is wasted if it is never used by latency-critical services. In the rest of the paper, we set this value to 2 if not otherwise specified, which balances between memory allocation speed and memory wastage. 
We also set the minimum amount of memory $min\_rsv$ that should be reserved after each execution of the management thread even if there is no newly incoming memory request. It allows that a burst of memory requests after an idle period can be quickly served. The value depends on the characteristics of latency-critical services. Empirically, we set this value to 5 MB. We use \texttt{mlock} system call to delegate virtual-physical mapping construction to kernel space.

There are two choices to implement the virtual-physical mapping construction function, 1) iterating through the allocated virtual memory addresses and filling them with `0', and 2) using the \texttt{mlock} system call to delegate the construction to the kernel space. We choose the second one for two reasons. First, our experiments find that using \texttt{mlock} system call is at least 40\% faster than the iteration approach for both heap memory and mmapped memory. Second, the \texttt{mlock} system call guarantees newly reserved physical memory not to be swapped into disks, which further accelerates memory allocation. After a chunk of reserved memory is allocated to a process, the \texttt{munlock} system call will be called on that address space to allow swapping on the chunk.

\begin{figure*}
\centering
\subfloat[Dedicated system.]{
\includegraphics[width=0.235\textwidth]{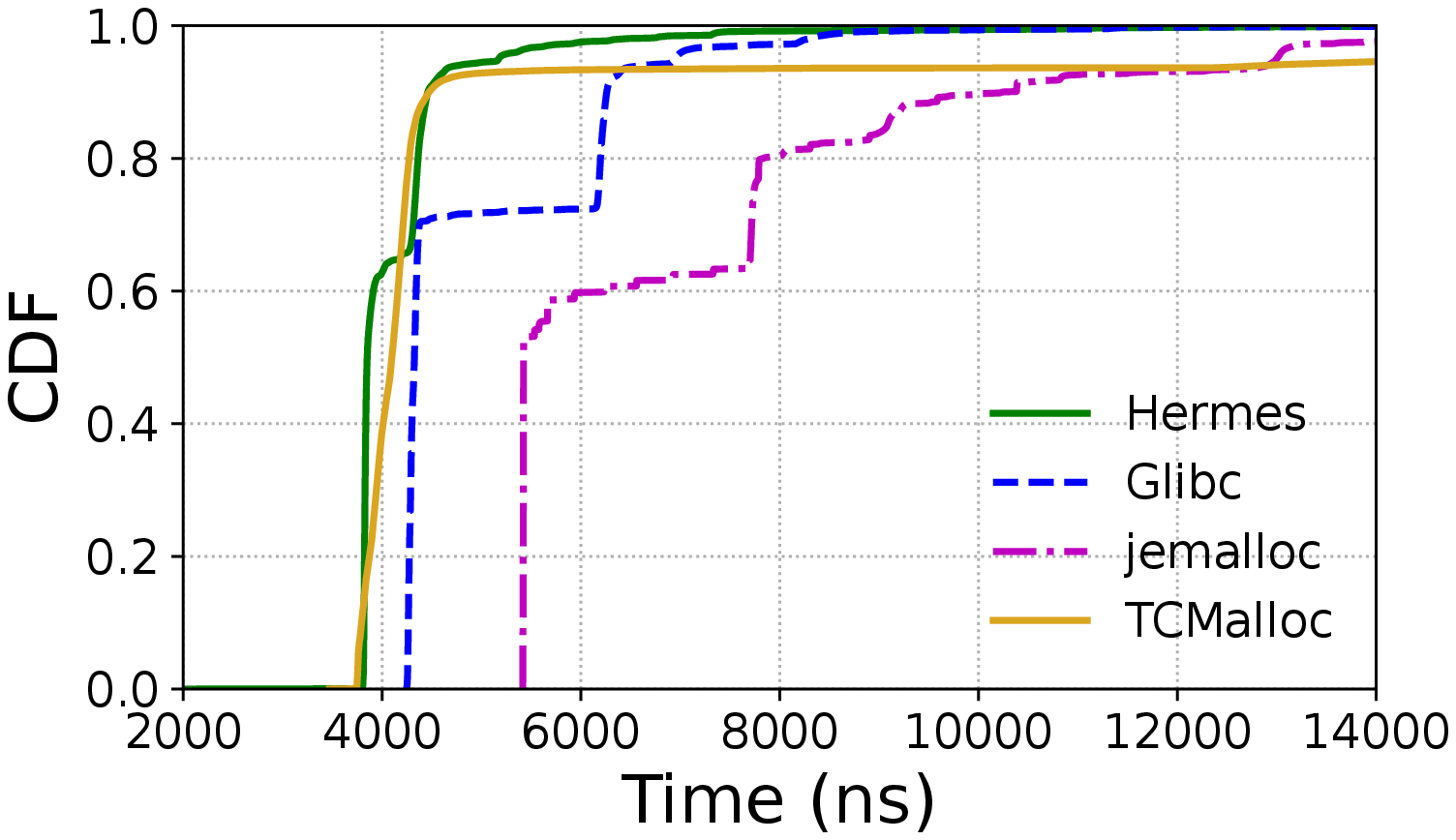}
\label{fig:micro-small-idle}
}
\subfloat[Anonymous pages pressure.]{
\includegraphics[width=0.235\textwidth]{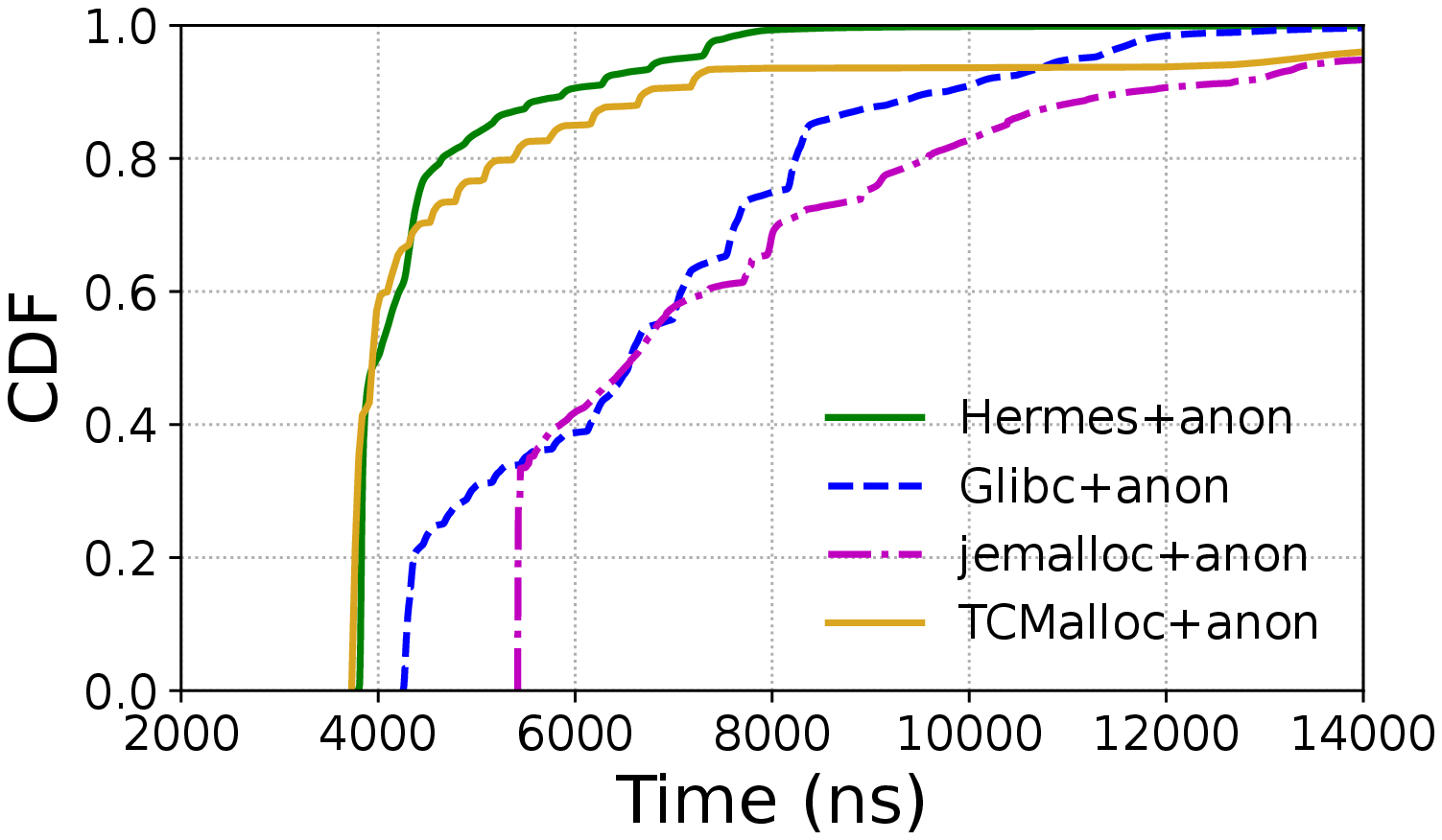}
\label{fig:micro-small-anon}
}
\subfloat[File cache pressure.]{
\includegraphics[width=0.235\textwidth]{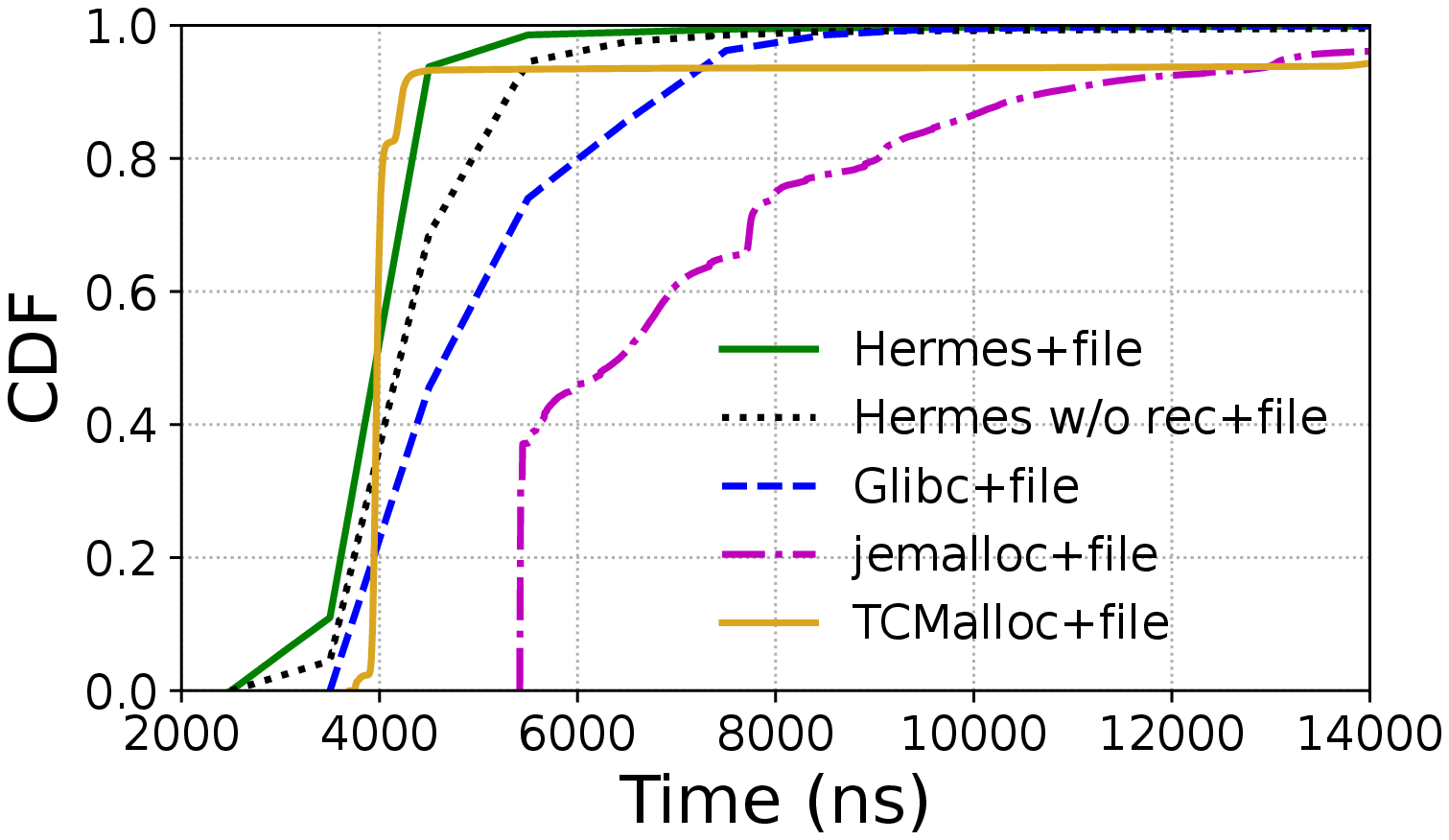}
\label{fig:micro-small-file}
}
\subfloat[Latency reduction by Hermes.]{
\includegraphics[width=0.235\textwidth]{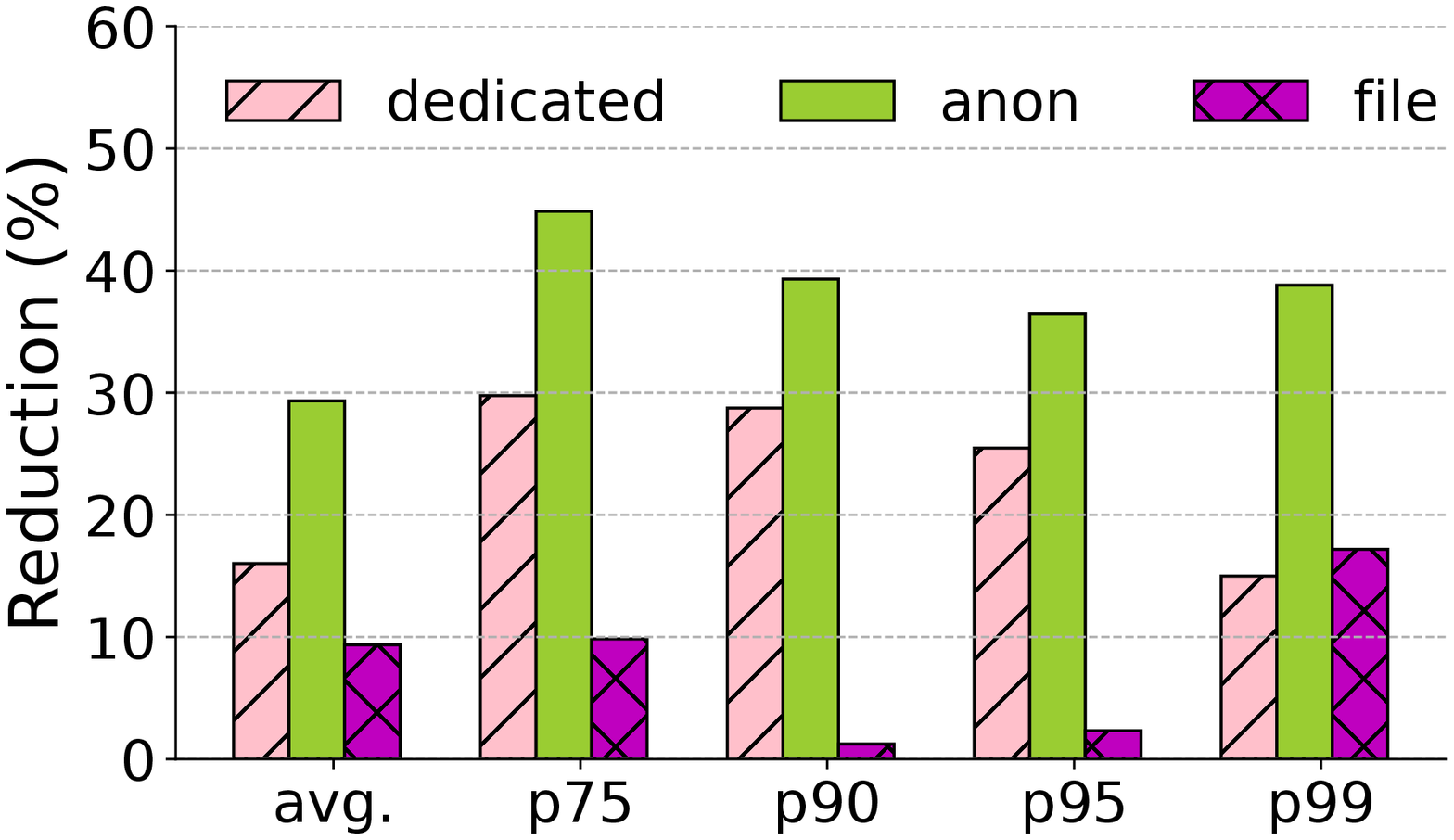}
\label{fig:small-reduction-hdd}
}
\caption{The memory allocation latency for small (1KB-size) memory requests on the HDD.}
\label{fig:micro-small}
\end{figure*}

\begin{figure*}
\centering
\subfloat[Dedicated system.]{
\includegraphics[width=0.235\textwidth]{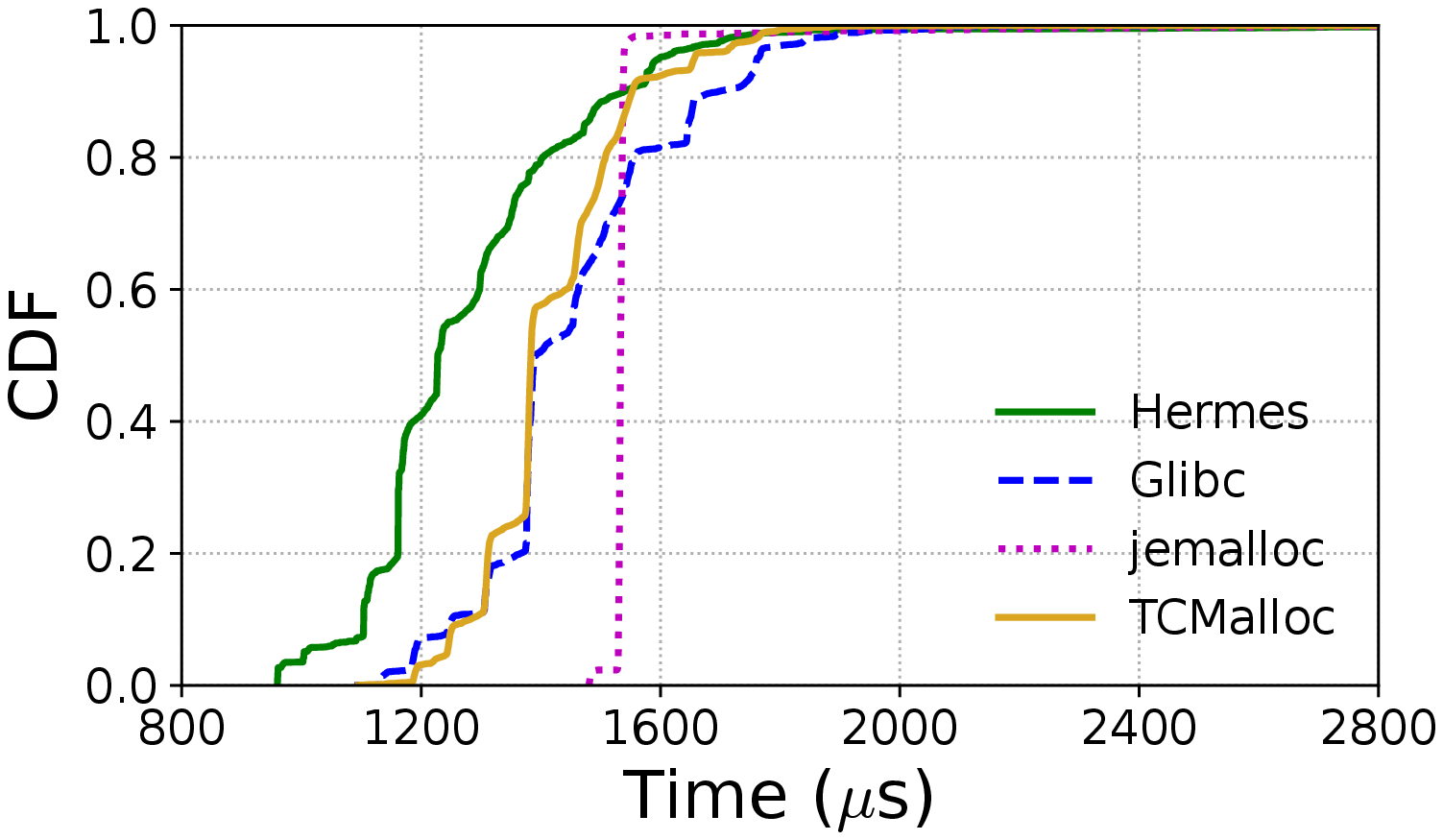}
\label{fig:micro-large-idle}
}
\subfloat[Anonymous pages pressure.]{
\includegraphics[width=0.235\textwidth]{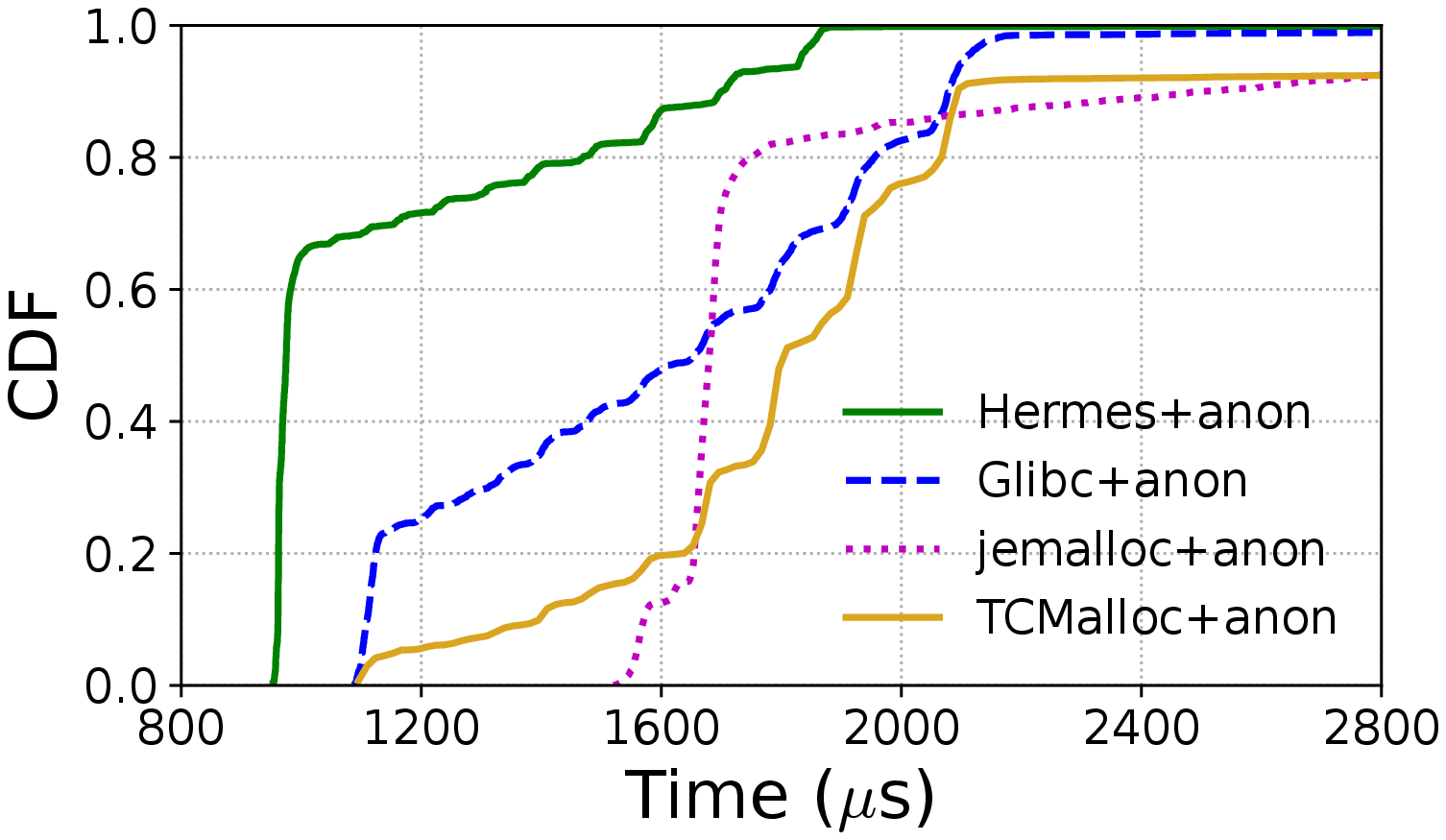}
\label{fig:micro-large-anon}
}
\subfloat[File cache pressure.]{
\includegraphics[width=0.235\textwidth]{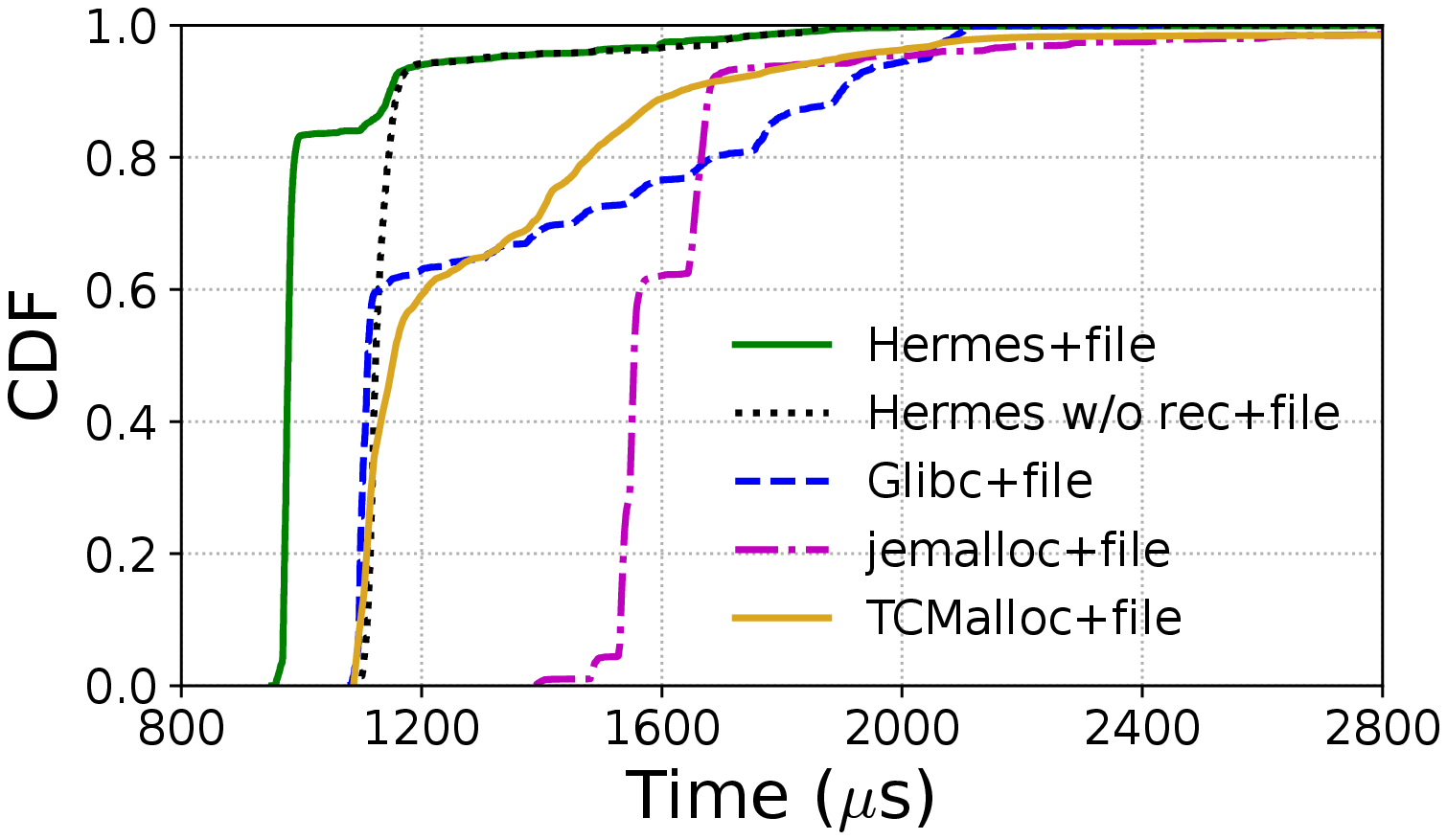}
\label{fig:micro-large-file}
}
\subfloat[Latency reduction by Hermes.]{
\includegraphics[width=0.235\textwidth]{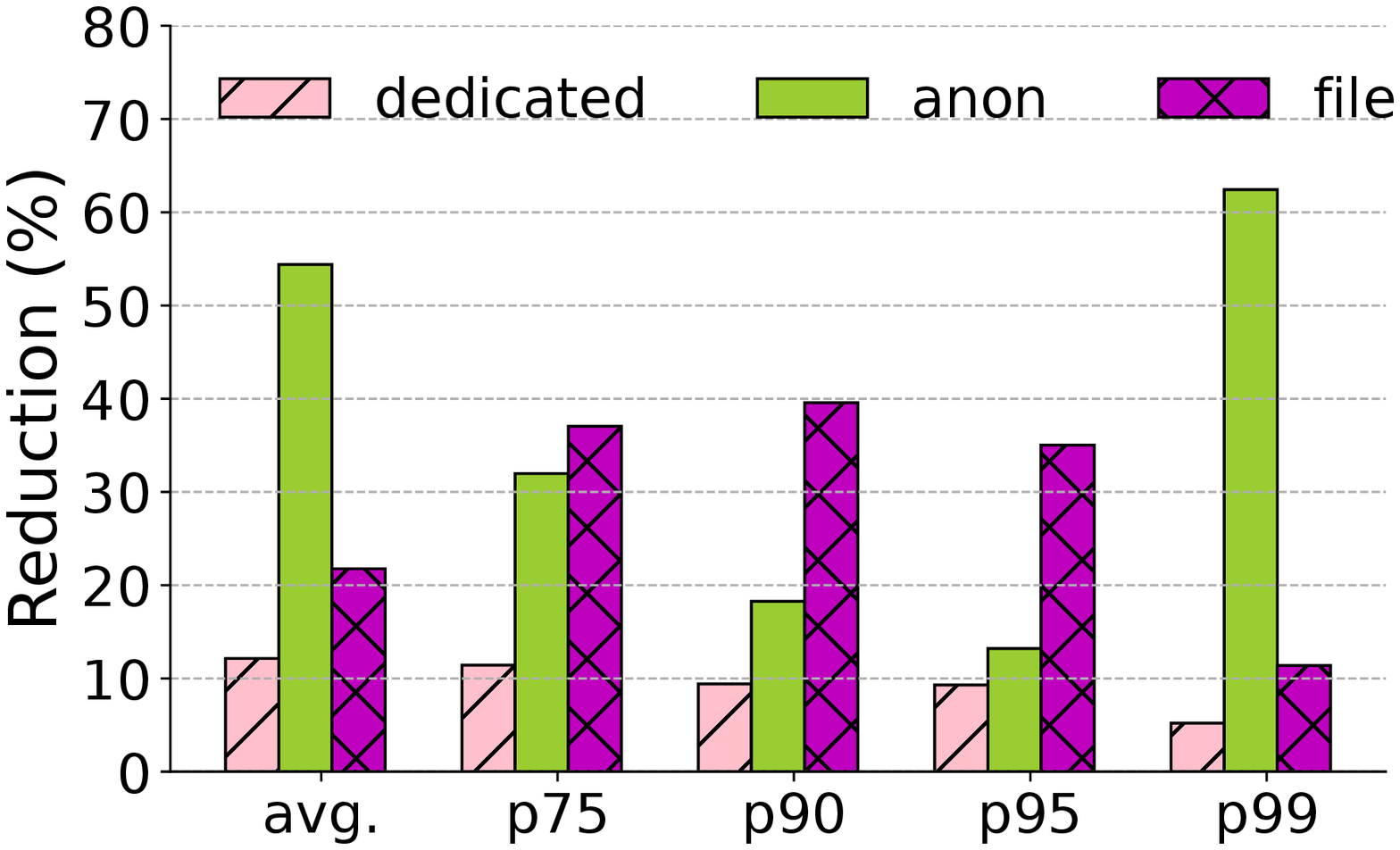}
\label{fig:large-reduction-hdd}
}
\caption{The memory allocation latency for large (256KB-size) memory requests on the HDD.}
\label{fig:micro-large}
\end{figure*}

The memory monitor daemon takes about 500 lines of C code. It is responsible for bookkeeping latency-critical services and batch jobs, and advising Linux OS to release file cache pages. It communicates with the modified Glibc with a shared memory area. Specifically, it uses the shared memory to store all the process IDs of latency-critical services specified by a system administrator. With the modified Glibc, a process examines whether its process ID is in the shared memory. If so, the modified Glibc initializes the memory management thread. When a process is no longer a latency-critical service, the administrator can simply remove its process ID. The monitor daemon keeps track of the data files loaded by batch jobs by  calling the \texttt{lsof} command. It uses the C library call \texttt{posix\_fadvise()} to release file cache pages, which is a wrapper function of the underlying system call \texttt{fadvise64()}. 
Hermes then adopts the default memory management in Glibc for this process. 

Hermes will be open sourced in Github.

\section{Evaluation}
\label{sec:evaluation}

\subsection{Evaluation Setup}
\label{sec:evaluation_setup}

We use both a micro benchmark and real-world latency-critical services to evaluate the performance of Hermes, and compare it to Glibc, jemalloc~\cite{Evans-BSDCan06-jemalloc}, and TCMalloc~\cite{tcmalloc}. Glibc is the most popular memory allocator in C/C++. Jemalloc is the default memory allocator for Redis~\cite{Redis}. TCMalloc is Google's customized implementation of \texttt{malloc()} function. All experiments are executed on a server that has two 2.4 GHz 8-core Intel Xeon E5-2630 CPUs, 128 GB DRAM, and 2 TB 7200 rpm HDD disks. The server is installed with Ubuntu 16.04 with Linux kernel-4.4.0. For all experiments, we pin latency-critical services and background processes onto different cores to avoid CPU interference.

\noindent
\textbf{Micro benchmark.} We implement a micro benchmark in C, which continuously calls \texttt{malloc} function to request memory until the total amount of requested memory reaches a specified threshold. We run the experiments in two settings referred as dedicated system and memory pressure. For the dedicated system setting, we run the micro benchmark alone on the nodes with sufficient memory. For the memory pressure setting, we generate the memory pressure for the micro benchmark by loading the node with either anonymous pages or file cache pages. We measure the memory allocation latency due to the three approaches. 

\noindent
\textbf{Real-world services.} We evaluate Redis~\cite{Redis} and Rocksdb~\cite{Rocksdb} as real-world latency-critical services under different memory pressure levels in Section~\ref{sec:real-world}. We measure three metrics in the experiments: 1) query latency of latency-critical services, 2) SLO violation of latency-critical services, and 3) throughput of batch job. To generate different levels of memory pressure, we configure the maximum logically available memory of batch jobs to 50\%, 75\%, 100\%, 125\% and 150\% of the memory capacity of the node. For example, on a node with 128 GB DRAM, 150\% memory pressure level suggests batch jobs can oversubscribe 192 GB (128 GB $\times$ 1.5) of DRAM. 

\noindent
\textbf{Parameter sensitivity.} We conduct experiments to evaluate parameter sensitivity in Section~\ref{sec:sensitivity}. Specifically, we run the micro benchmark and evaluate its latency under different values of reservation factor $RSV\_FACTOR$. 


\subsection{Micro Benchmark}
\label{sec:micro-hdd}
We evaluate the performance of Hermes under three scenarios: a dedicated system with sufficient memory, anonymous page pressure, and file cache pressure. Under file cache pressure, we also show the performance of Hermes when it is disabled with proactive reclamation, denoted as ``Hermes w/o rec',  to demonstrate the performance gain due to proactive reclamation. The anonymous page pressure is made by a process that keeps allocating memory until the system available memory drops below 300 MB. The file cache pressure is made by a process that repeatedly reads 10 GB files and occupies the rest of the system memory with anonymous pages. We develop the micro benchmark by continuously sending fix-sized memory requests until the total requested memory reaches 1 GB. We use 1KB-size and 256KB-size memory requests to evaluate the allocation latency of heap memory and mmapped memory.

Figure~\ref{fig:micro-small}(a)-(c) and Figure~\ref{fig:micro-large}(a)-(c) show the CDFs of memory allocation latency of 1KB-size and 256KB-size requests under a dedicated system, anonymous page pressure (``+anon'' suffix), and file cache pressure (``+file'' suffix), respectively. For small memory requests, Hermes achieves the lowest latency at every percentile compared to Glibc and jemalloc in all three cases. TCMalloc presents low latency on average. However, it has very high tail latency in all three cases.  As for large memory requests, jemalloc presents longer but more stable latency under a dedicated system. However, Hermes outperforms both Glibc, jemalloc and TCMalloc when the system is under memory pressure. Jemalloc and TCMalloc presents very long tail latency under memory pressure. Specifically, we show the latency reduction of Hermes at each percentile compared to Glibc in Figures~\ref{fig:small-reduction-hdd} and \ref{fig:large-reduction-hdd} since Glibc outperforms jemalloc in most cases. For 1KB-size requests, Hermes reduces the average latency by 16.0\%, 29.3\%, 9.4\%, and the $99^{th}$ percentile latency by 15.0\%, 38.8\%, 17.2\% in the three scenarios, respectively. For 256KB-size requests, Hermes reduces the average latency by 12.1\%, 54.4\%, 21.7\%, and the $99^{th}$ percentile latency by 5.2\%, 62.4\%, 11.4\%, respectively. Hermes outperforms the default Glibc at each percentile in all scenarios. The allocation latency is as low as $4 \mu s$ for small requests and $1 ms$ for large requests.

By comparing the ``dedicated'' and ``file'' bars in Figure~\ref{fig:small-reduction-hdd} to those in Figure~\ref{fig:large-reduction-hdd}, the performance gain by Hermes under a dedicated system and under file cache pressure for large requests is more significant than that for small requests. The reason is that large requests take a long time to be allocated in the default Glibc. 
By comparing the ``anon'' bar to the ``dedicated'' and ``file'' bars in Figure~\ref{fig:small-reduction-hdd} or Figure~\ref{fig:large-reduction-hdd}, we observe that Hermes generally achieves more performance improvement under anonymous pressure for both small and large requests compared to those under file cache pressure. The reason is that it is faster to reclaim file cache pages in the default Linux kernel since unmodified file cache pages are directly released without I/O operations. For anonymous pages, however, each of them must be swapped into disks before released, causing much longer delay due to I/O operations. 

\noindent
\textbf{Proactive reclamation.} Figures~\ref{fig:micro-small-file} and \ref{fig:micro-large-file} show that ``Hermes w/o paging'' achieves similar memory allocation latency at low percentiles compared with the default Glibc, but it significantly reduces the latency at high percentiles. Full Hermes further improves the average latency over ``Hermes w/o paging''. 

\subsection{Two Real-world Latency-critical Services}
\label{sec:real-world}

\begin{figure}
\subfloat[Small requests]{
\includegraphics[width=0.48\linewidth]{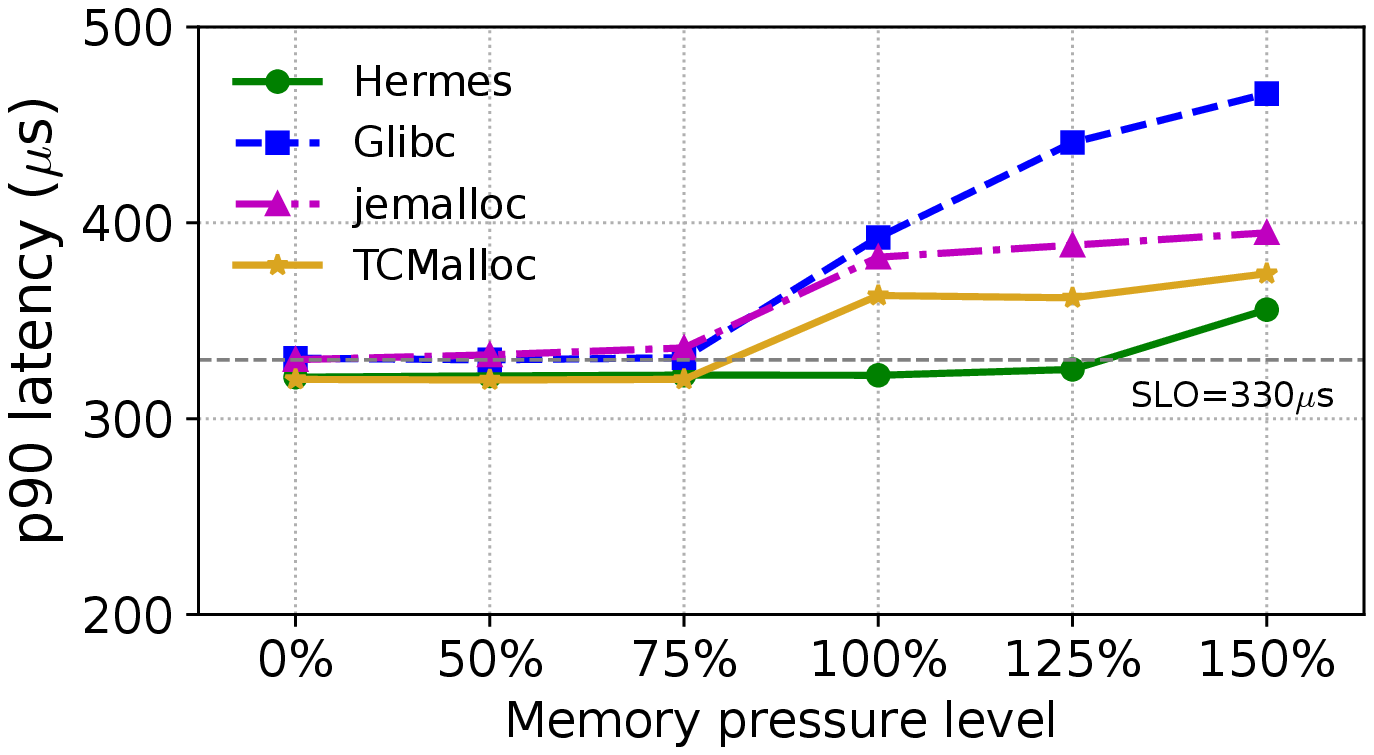}
\label{fig:redis-pressure-small}
}
\subfloat[Large requests]{
\includegraphics[width=0.48\linewidth]{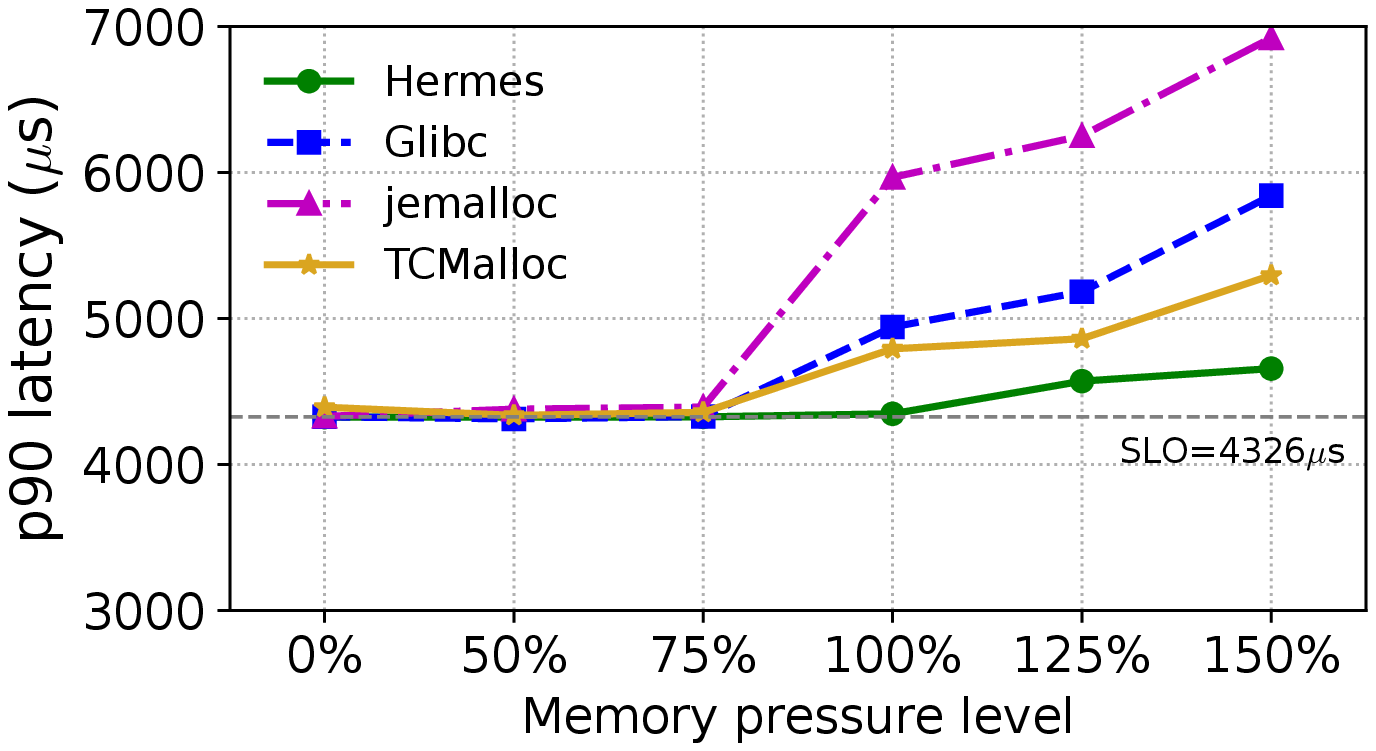}
\label{fig:large-pressure-large}
}
\caption{The $90^{th}$ percentile query latency of Redis.}
\label{fig:redis-pressure}
\end{figure}

\subsubsection{Query latency and SLOs}

We evaluate the query latency reduction on real-world latency-critical services by Hermes compared to Glibc, jemalloc and TCMalloc under different memory pressure. We use Redis-5.0.5~\cite{Redis} and Rocksdb-6.4.0~\cite{Rocksdb} as two representative real-world services. Redis is an in-memory key-value store for fast data access. Rocksdb is a disk-based persistent key-value store for fast storage environments. It uses memory as data cache. These services are usually used for intermediate or temporary data storage. Thus, they frequently allocate and release memory. For both Redis and Rocksdb, we implement a program to continuously generate requests. One request consists of one insertion operation followed by one read operation. We use 1KB-size and 200KB-size data records referenced as small and large memory requests, respectively. For each data insertion execution, we insert the data until it reaches 2 GB. To inject memory pressure, we run Spark Kmeans and Spark PageRank as batch jobs on the host node. The jobs are from HiBench-6.0~\cite{huang2010hibench} using its default huge data size. We run Spark-2.3.0 on Hadoop-2.7.3~\cite{vavilapalli2013apache}.

\begin{figure}
\subfloat[Small requests]{
\includegraphics[width=0.48\linewidth]{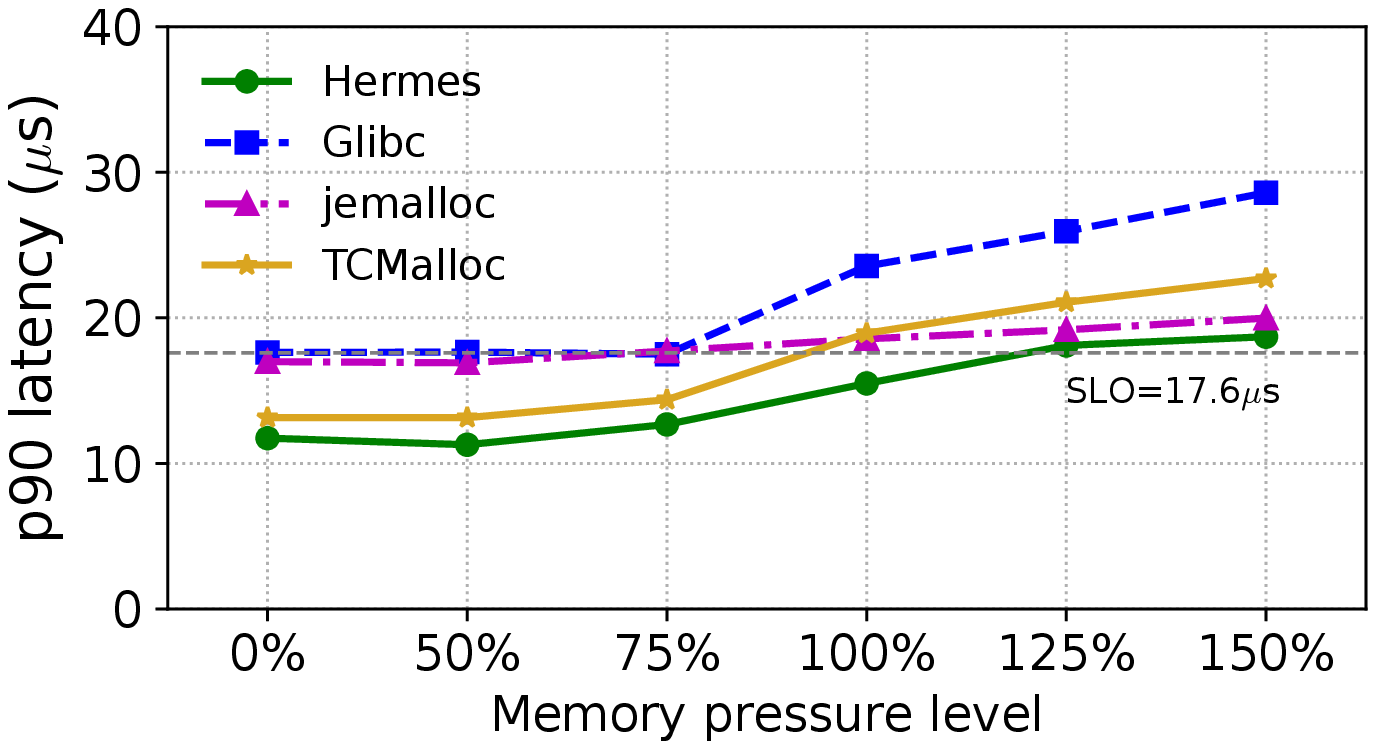}
\label{fig:rocksdb-pressure-small}
}
\subfloat[Large requests]{
\includegraphics[width=0.48\linewidth]{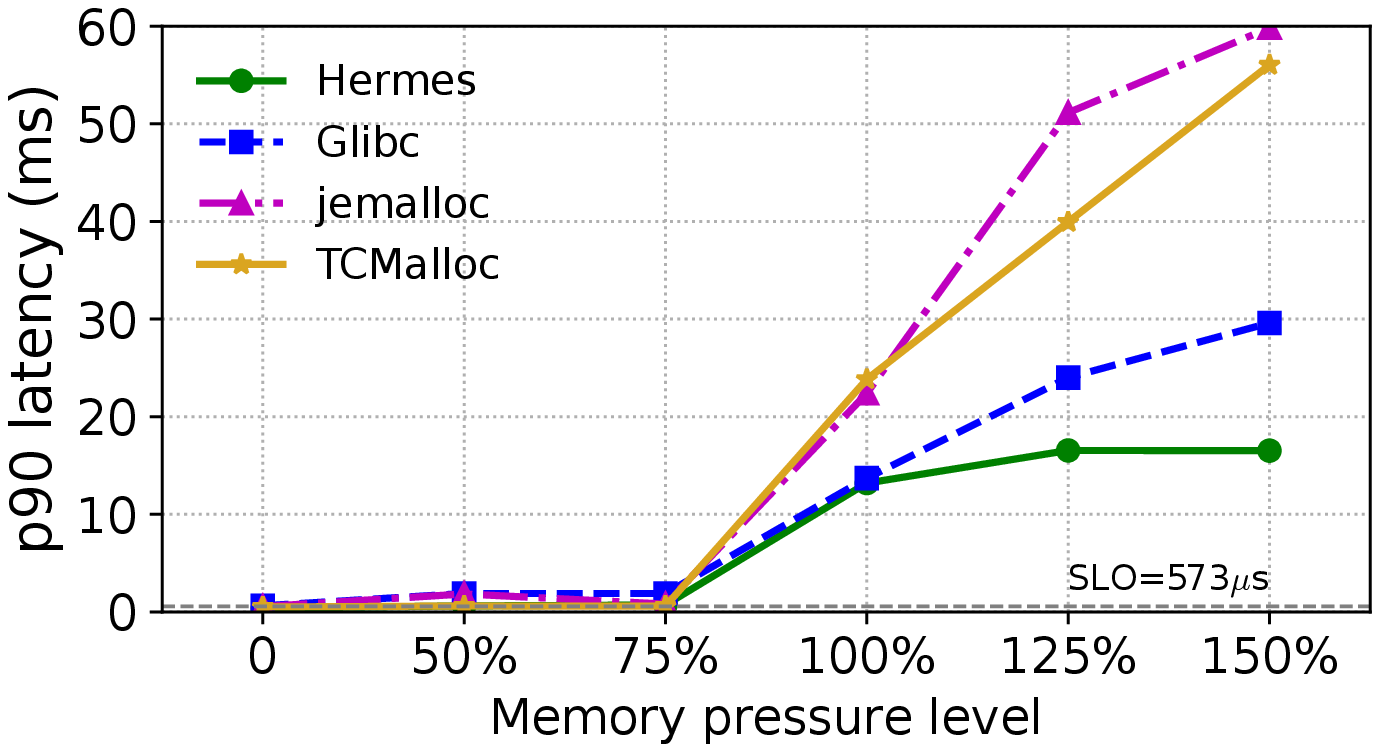}
\label{fig:rocksdb-pressure-large}
}
\caption{The $90^{th}$ percentile query latency of Rocksdb.}
\label{fig:rocksdb-pressure}
\end{figure}

Since there is not a magic value to define the SLO of each service, we adopt the $90^{th}$ percentile latency by the default Glibc under a dedicated system (w/o memory pressure) as the SLO, which is a rather strict value. The rational is that latency-critical services like web search commonly distribute requests across many servers. The end-to-end response time is determined by the slowest individual latency~\cite{TailAtScale,Zhu-SOCC17-WorkloadCompactor,Berger-OSDI18-RobinHood,fried2020caladan}. Thus, the $90^{th}$ percentile latency is a critical metric in measuring the SLO of latency-critical services.

\noindent \textbf{Latency reduction.} Figures~\ref{fig:redis-pressure} and~\ref{fig:rocksdb-pressure} show the $90^{th}$ percentile query latency under different memory pressure levels for Redis and Rocksdb, respectively. The horizontal dash line represents the target SLO in each situation. In Redis, the SLOs are $330 \mu s$ and $4,326 \mu s$ for small and large requests, respectively. In Rocksdb, the SLOs are $17 \mu s$ and $573 \mu s$ for small and large requests, respectively. 

The results show that Hermes outperforms Glibc, jemalloc and TCMalloc in reducing the $90^{th}$ percentile query latency in all scenarios for both Redis and Rocksdb. Specifically, with a dedicated system (0\% memory pressure) or a low memory pressure level (50\% and 75\%), Hermes achieves similar or slightly lower $90^{th}$ percentile latency compared to Glibc, jemalloc and TCMalloc. 
With a moderate memory pressure level (100\% and 125\%), Hermes can meet the SLO targets for small requests while Glibc, jemalloc and TCMalloc incur significant SLO violation. With a severe memory pressure level ($>$ 125\%), all three approaches incur non-trivial SLO violation but Hermes significantly outperforms the others. We observe that large requests in Rocksdb under high memory pressure experience tens of milliseconds of latency. 

\begin{figure}
\subfloat[Small requests w/ batch jobs]{
\includegraphics[width=0.48\linewidth]{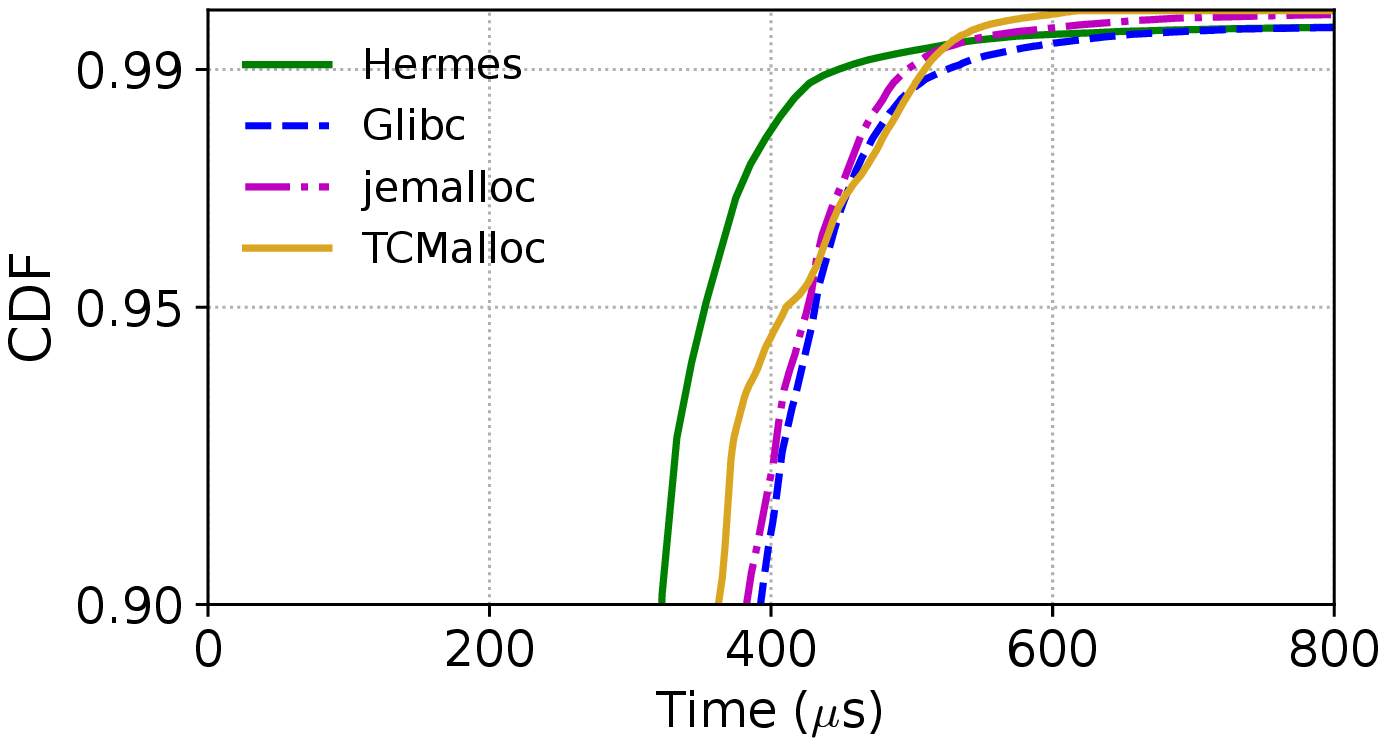}
\label{fig:redis-cdf-small}
}
\subfloat[Large requests w/ batch jobs]{
\includegraphics[width=0.48\linewidth]{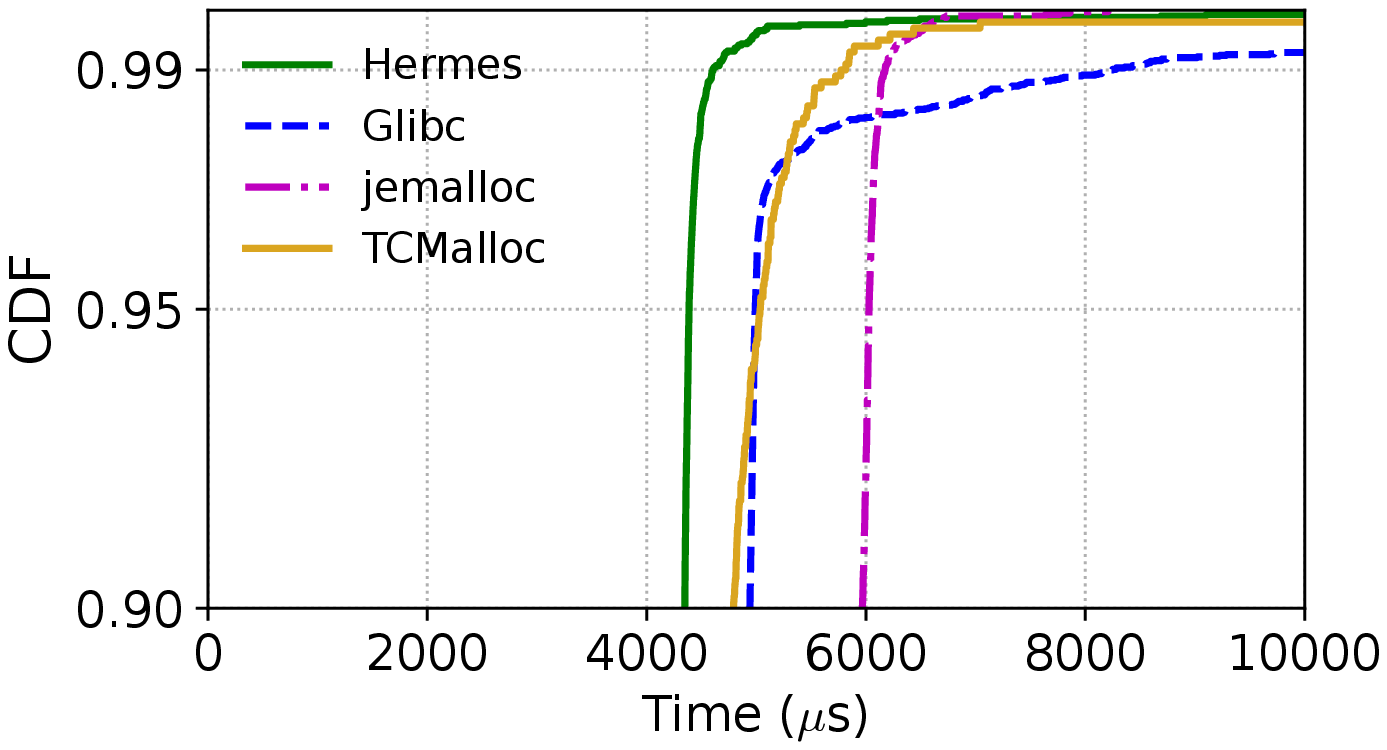}
\label{fig:redis-cdf-large}
}
\caption{Latency of Redis under 100\% memory pressure.}
\label{fig:redis-cdf}
\end{figure}

\begin{figure}
\subfloat[Small requests w/ batch jobs]{
\includegraphics[width=0.48\linewidth]{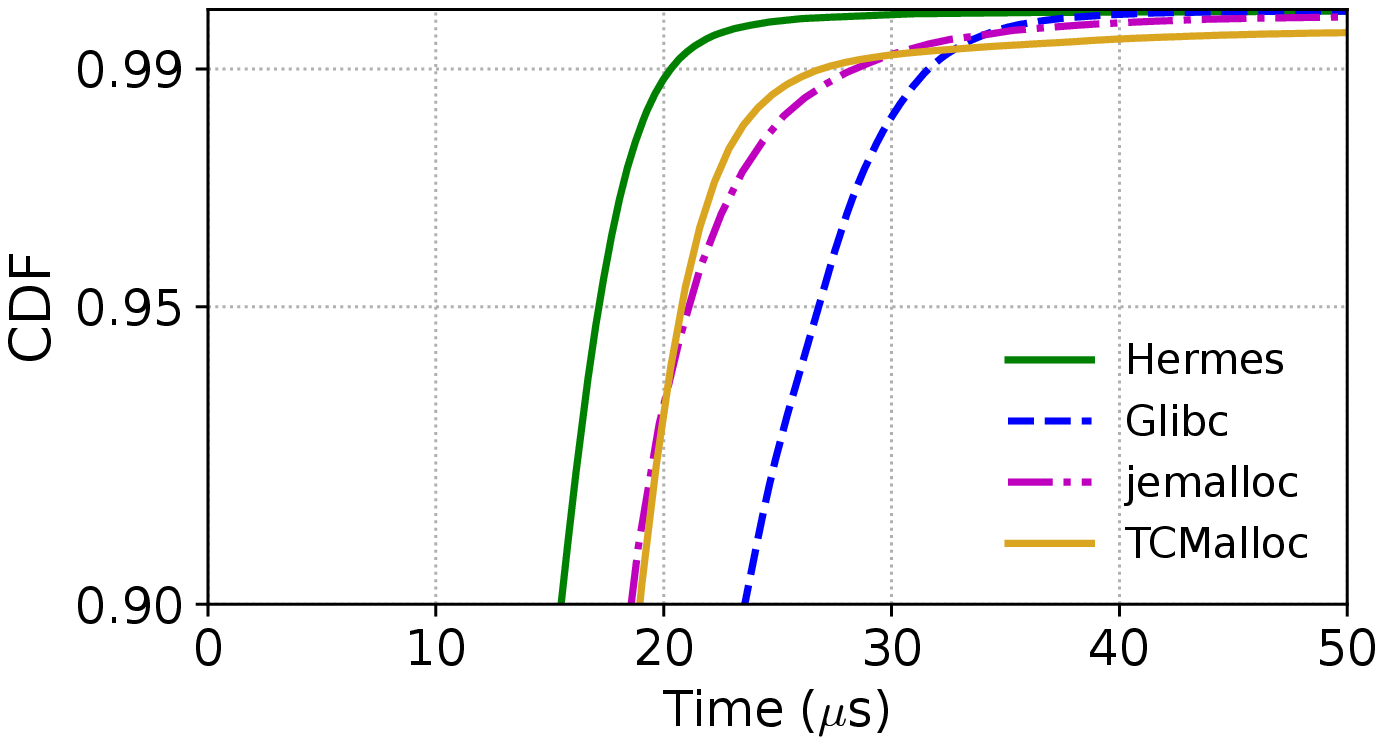}
\label{fig:rocksdb-cdf-small}
}
\subfloat[Large requests w/ batch jobs]{
\includegraphics[width=0.48\linewidth]{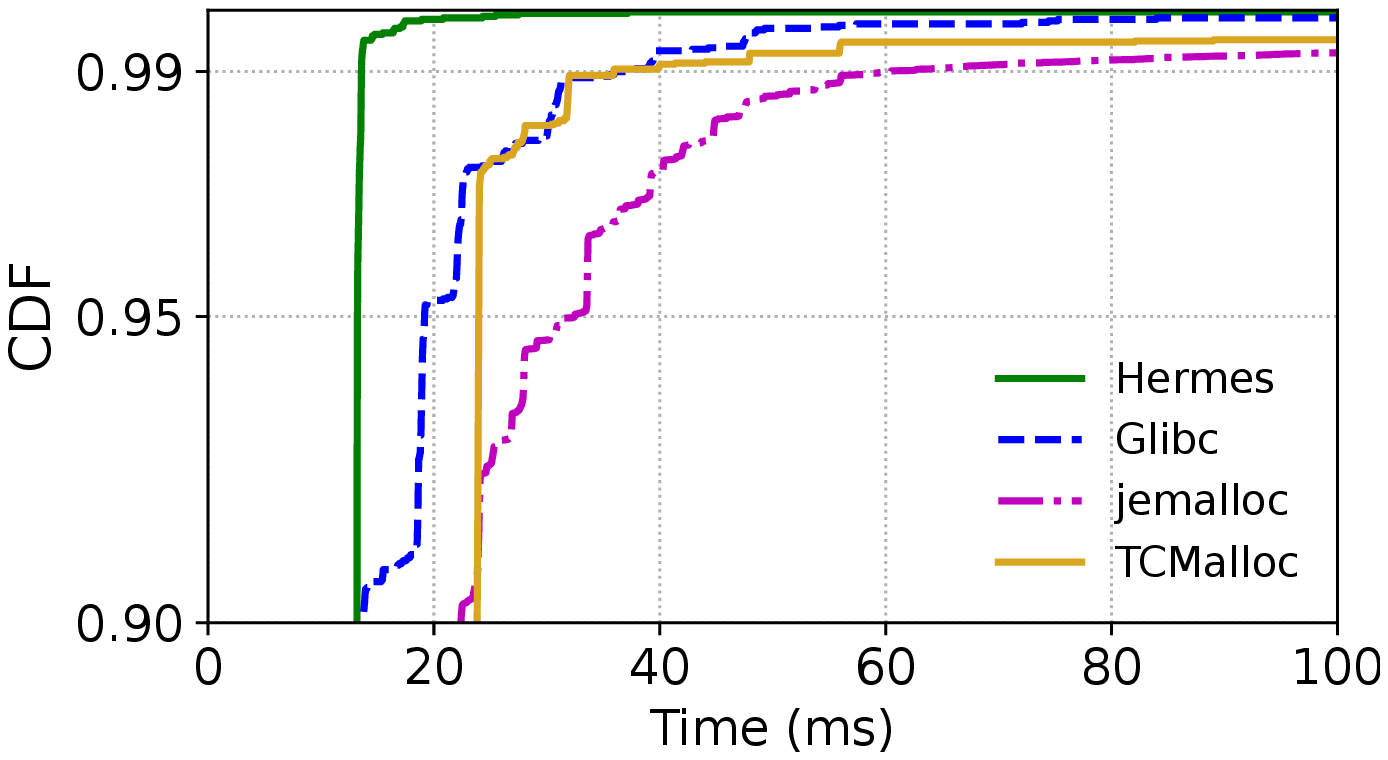}
\label{fig:rocksdb-cdf-large}
}
\caption{Rocksdb latency under 100\% memory pressure.}
\label{fig:rocksdb-cdf}
\end{figure}

Under job co-location, severe memory pressure is usually addressed by a system administrator while memory pressure around the 100\% level is more likely to happen due to the dynamic memory consumption of latency-critical services and batch jobs. Thus, we plot the CDF of the query latency under such a scenario for Redis and Rocksdb in Figures~\ref{fig:redis-cdf} and \ref{fig:rocksdb-cdf}, respectively. Hermes achieves the lowest latency for both services. Compared to Glibc, it reduces the average ($99^{th}$ percentile) latency by up to 17.0\% (40.6\%) for Redis and 20.6\% (63.4\%) for Rocksdb. 

\begin{figure}
\subfloat[Small requests]{
\includegraphics[width=0.46\linewidth]{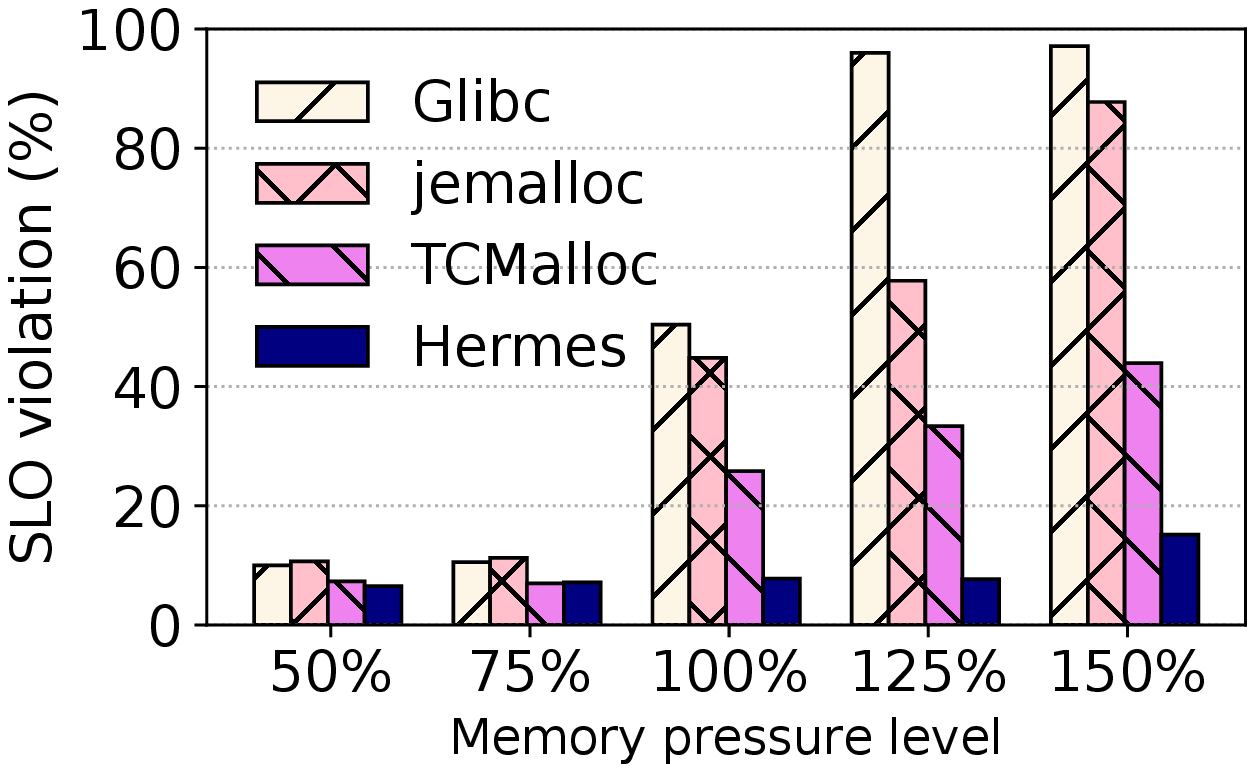}
\label{fig:redis-pressure-small-violation}
}
\subfloat[Large requests]{
\includegraphics[width=0.46\linewidth]{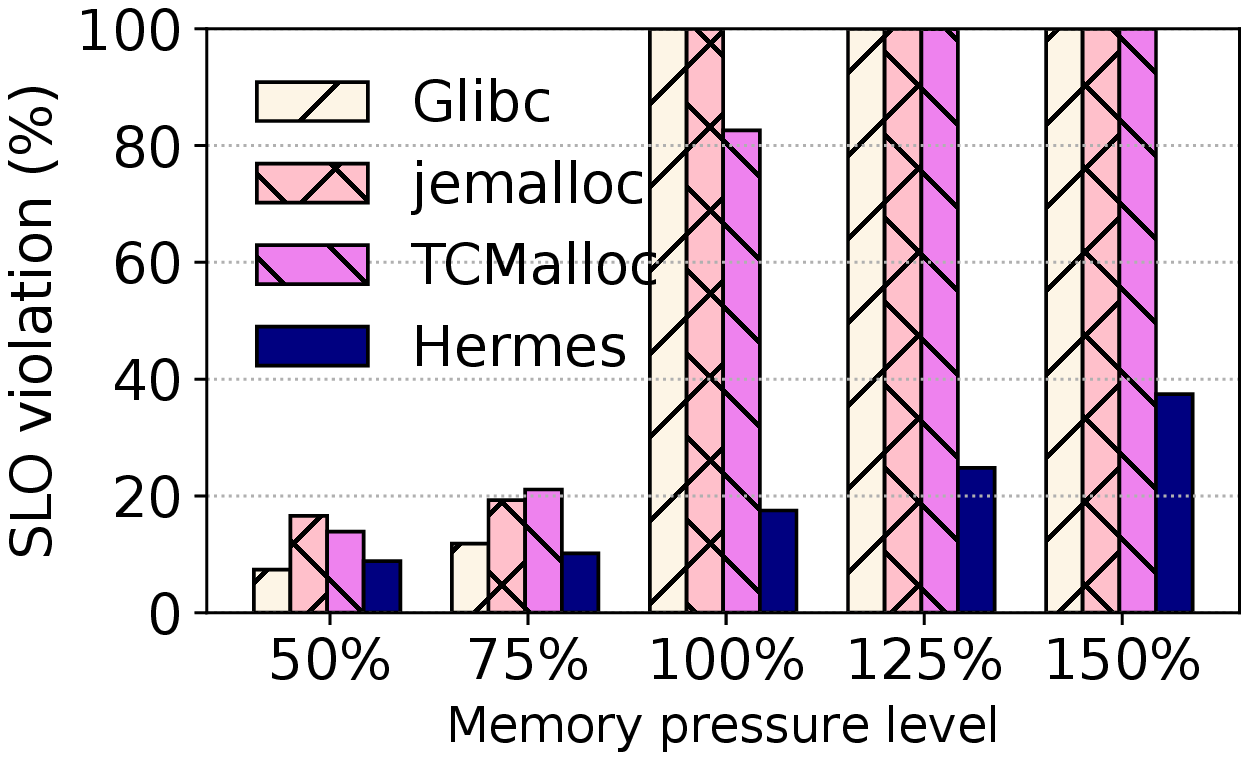}
\label{fig:redis-pressure-large-violation}
}
\caption{The SLO violation ratio of Redis requests.}
\label{fig:redis-pressure-violation}
\end{figure}

\begin{figure}
\subfloat[Small requests]{
\includegraphics[width=0.46\linewidth]{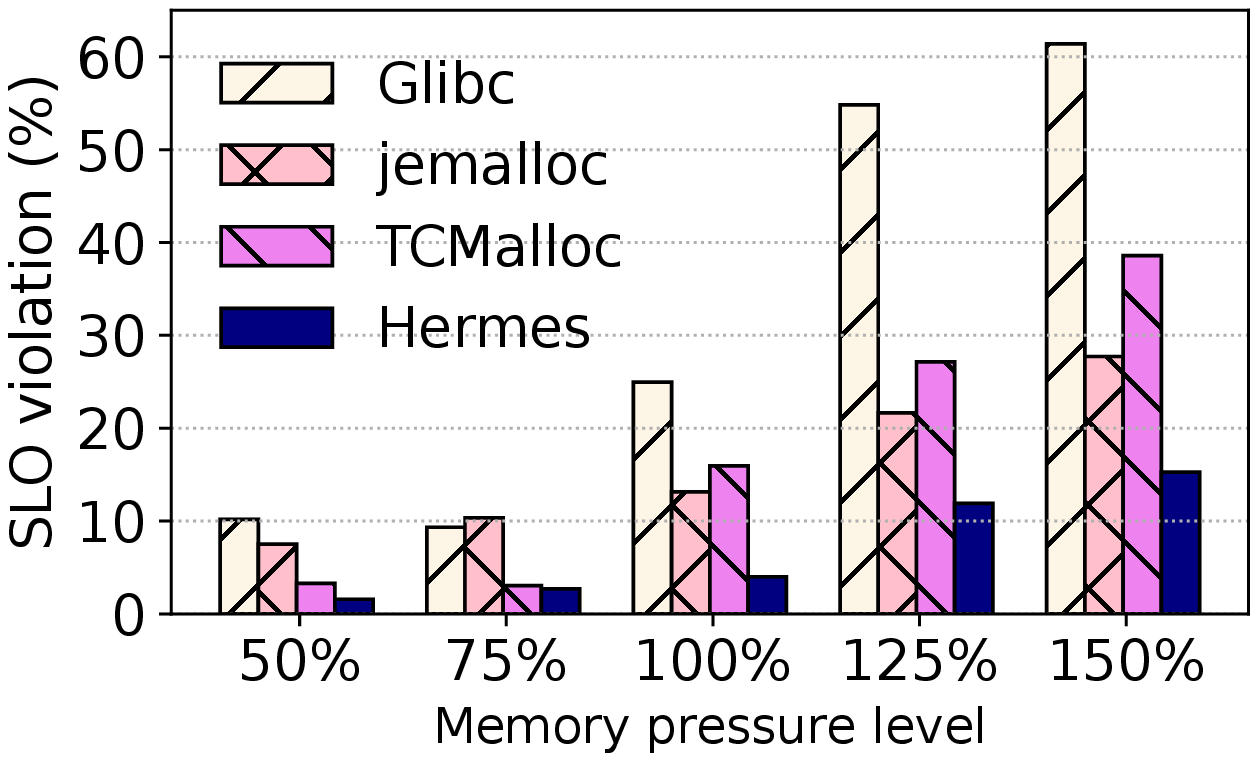}
\label{fig:rocksdb-pressure-small-violation}
}
\subfloat[Large requests]{
\includegraphics[width=0.46\linewidth]{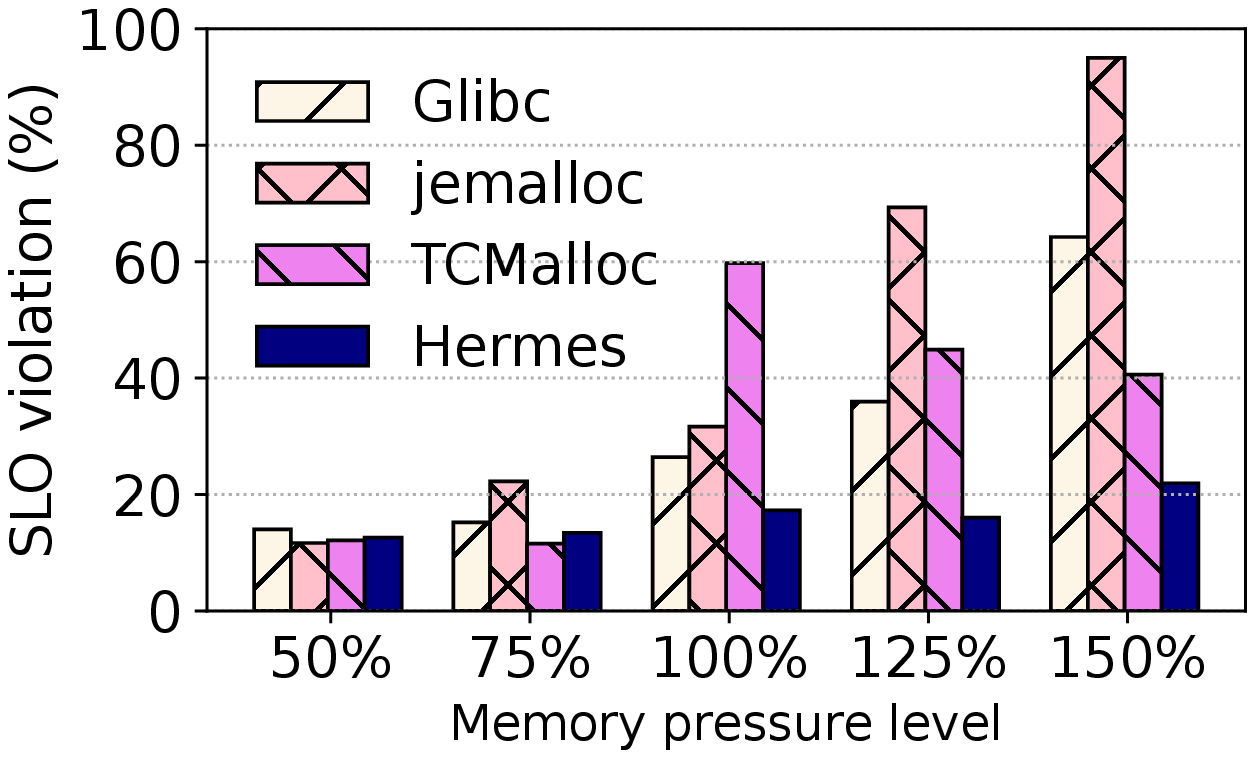}
\label{fig:rocksdb-pressure-large-violation}
}
\caption{The SLO violation ratio of Rocksdb requests.}
\label{fig:rocksdb-pressure-violation}
\end{figure}

\noindent
\textbf{SLO violation.} Figure~\ref{fig:redis-pressure-violation} and Figure~\ref{fig:rocksdb-pressure-violation} show the ratios of SLO violation with Hermes, Glibc, jemalloc and TCMalloc under different memory pressure levels for Redis and Rocksdb, respectively. For Redis, Hermes achieves the SLO violation ratio lower than 10\% under a low memory pressure level (i.e., 50\% and 75\%). The results for Rocksdb are similar. The reason is that Hermes builds the virtual-physical mapping in advance such that incoming memory requests can be immediately served. The most significant results are those under 100\% or higher memory pressure levels which usually happen in a multi-tenant system. Under such a memory pressure level, compared to the default Glibc, jemalloc, and TCMalloc, Hermes reduces the SLO violation of Redis by up to 83.6\%, and reduces the SLO violation of Rocksdb by up to 84.3\%.

\subsubsection{Batch job throughput}

We examine the throughput of batch jobs co-located with latency-critical services. We submit Spark Kmeans jobs and keep three concurrent job instances in the node. Each Kmeans job runs in eight Yarn containers and requests around 40GB memory. This generates the 100\% memory pressure level. We send data insertion, read, and deletion requests to the latency-critical services such that stored data size varies from 20GB to 40GB. The co-location experiment runs for 24 hours in each of the three scenarios: Default, Hermes, and Killing.

\begin{table}
\footnotesize
\caption{Throughput of batch jobs.}
\begin{tabular*}{\linewidth}{p{1cm}<{\centering} | p{1.4cm}<{\centering} | p{1.4cm}<{\centering} | p{1.4cm}<{\centering} | p{1.4cm}<{\centering}}
\hline
& \textbf{Default} & \textbf{Hermes} & \textbf{Killing} &  \textbf{Dedicated}\\
\hline
\textbf{Redis} & 212 & 194 & 123 & 0 \\
\hline
\textbf{Rocksdb} & 380 & 364 & 267 & 0\\
\hline
\end{tabular*}
\label{tbl:throughput}
\end{table}

\begin{itemize}
\item \textbf{Default.} We co-locate batch jobs and latency-critical services with the default GNU/Linux stack.

\item \textbf{Hermes.} We co-locate batch jobs and latency-critical services with Hermes.

\item \textbf{Killing.} Upon Default, we kill the latest launched container of a batch job when node memory is insufficient, which frees up memory. Killing the container results in the least progress loss of the batch job.
\end{itemize}

Table~\ref{tbl:throughput} gives the number of the finished batch jobs in the three co-location scenarios as well as in a dedicated system where is no throughput of batch jobs. Both Default and Hermes achieve much higher throughput than that of Killing. Hermes achieves slightly lower throughput to that of Default. In return, it significantly reduces the query latency and SLO violation of latency-critical services, the principle requirement of job co-location. We notice that the throughput of co-location with Rocksdb is higher than that of Redis. The reason is that Redis is a memory-based KV store that keeps all data in DRAM. Rocksdb is a disk-based KV store that has much lower memory consumption than Redis. Thus, more memory can be allocated to batch jobs. Experimental results find that job co-location due to Hermes renders about 98.5\% average node memory utilization. 


%

\textbf{Under a Dedicated System.} We evaluate the performance of real-world services Redis and Rocksdb by Hermes, Glibc and jemalloc under a dedicated system. Compared to Glibc and jemalloc, Hermes renders similar or slightly better average, $90^{th}$, and $99^{th}$ percentile query latency. 

\subsection{Parameter Sensitivity}
\label{sec:sensitivity}

\begin{figure}
\subfloat[Dedicated system]{
\includegraphics[width=0.47\linewidth]{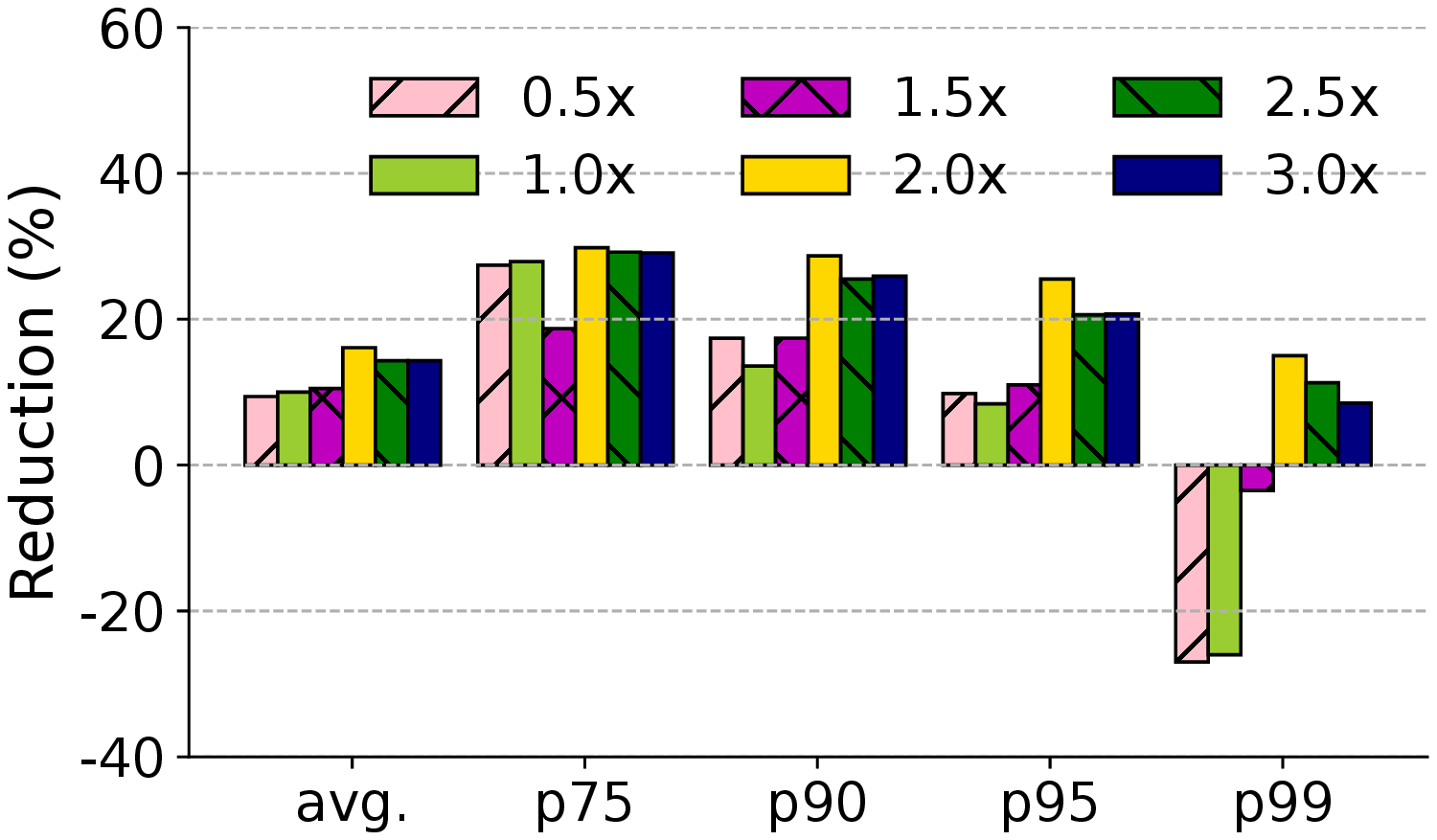}
\label{fig:small-sensitivity-idle}
}
\subfloat[Anonymous pressure]{
\includegraphics[width=0.47\linewidth]{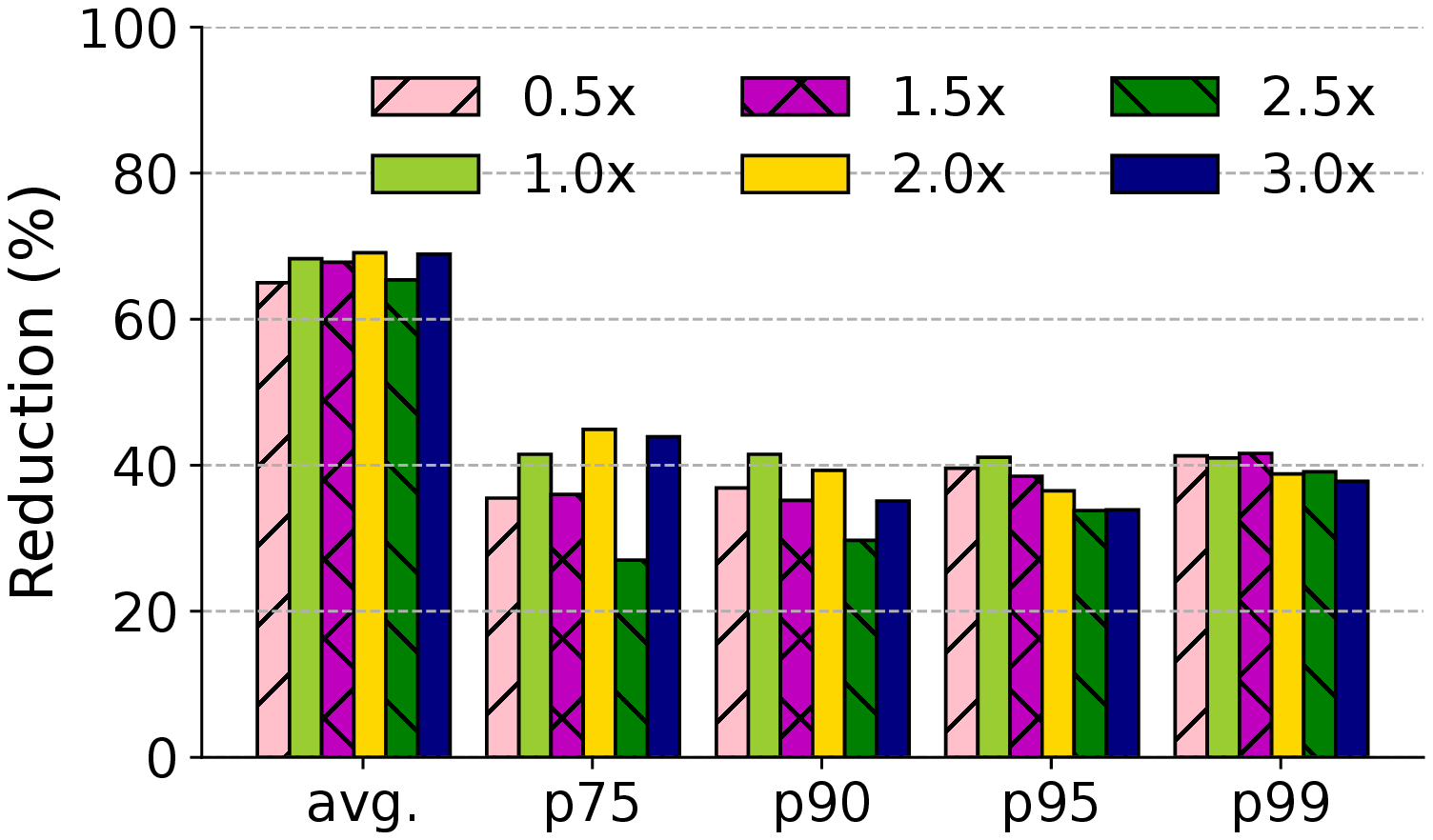}
\label{fig:small-sensitivity-anon}
}
\caption{Latency reduction for small requests.}
\label{fig:small-sensitivity}
\end{figure}

\begin{figure}
\subfloat[Dedicated system]{
\includegraphics[width=0.47\linewidth]{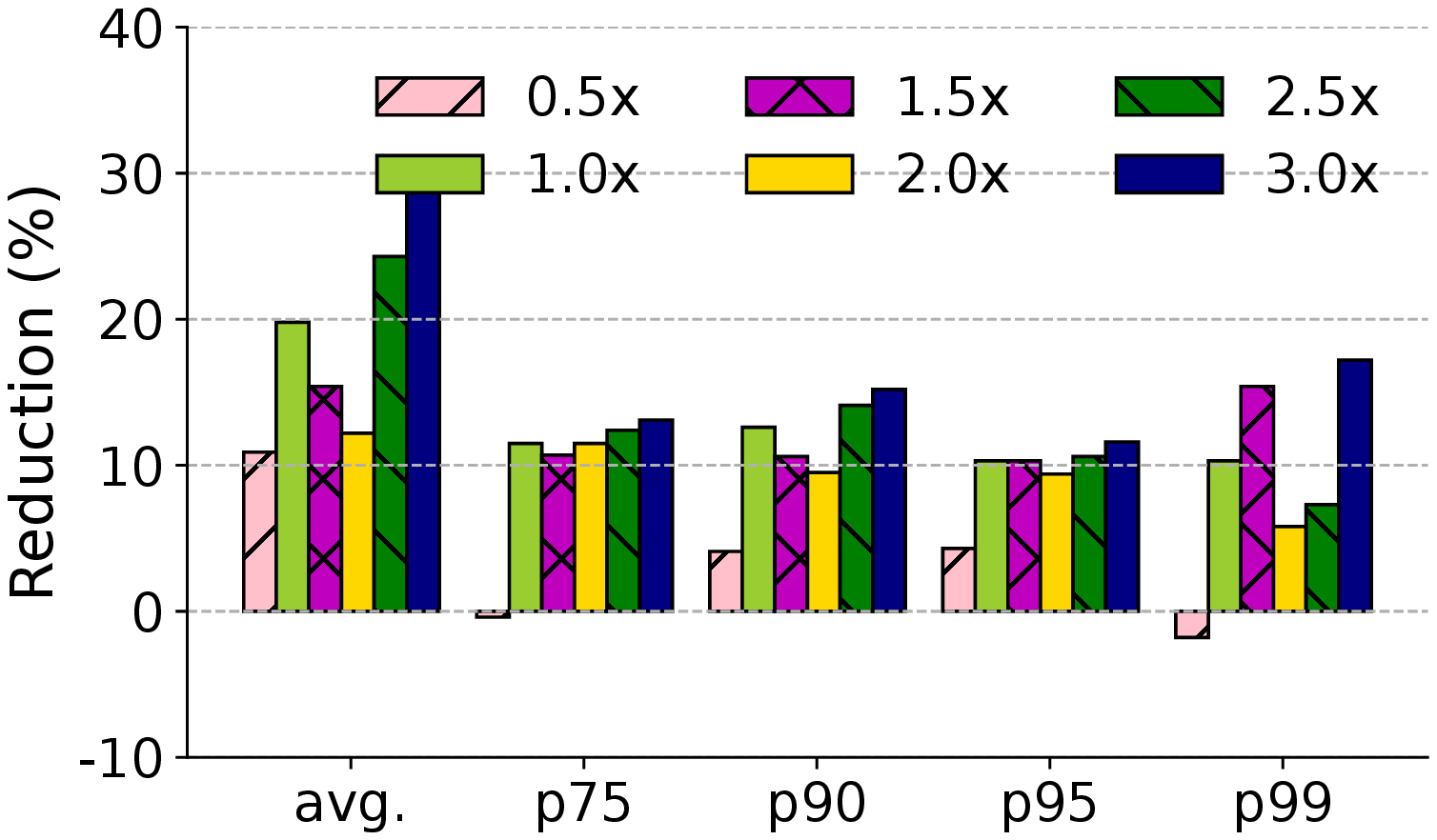}
\label{fig:large-sensitivity-idle}
}
\subfloat[Anonymous pressure]{
\includegraphics[width=0.47\linewidth]{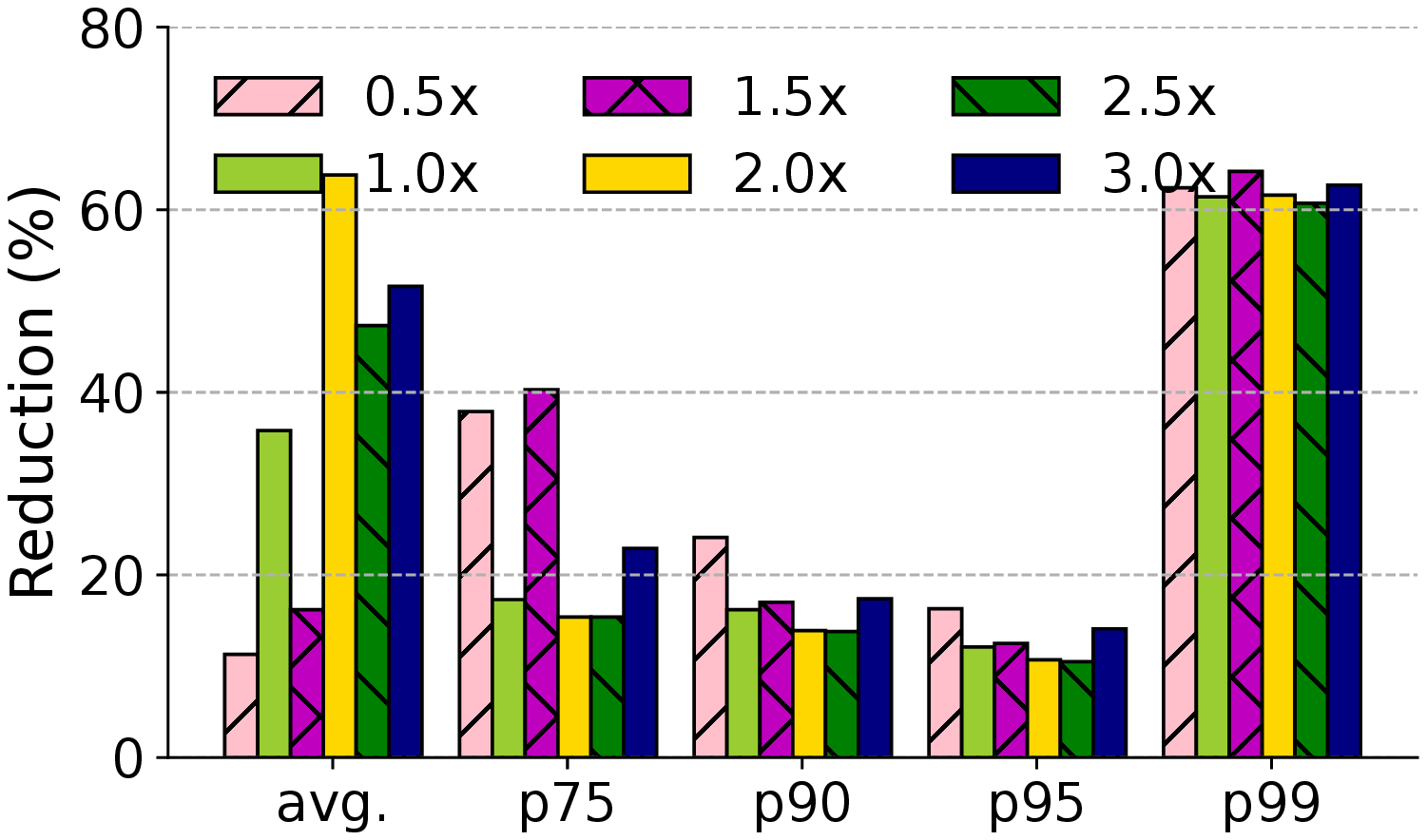}
\label{fig:large-sensitivity-anon}
}
\caption{Latency reduction for large requests.}
\label{fig:large-sensitivity}
\end{figure}

We evaluate the impact of parameter sensitivity. Specifically, we change the value of reservation factor $RSV\_FACTOR$ raging from 0.5 to 3, and evaluate the memory allocation latency under each value for both small and large memory requests using the micro benchmark. We run the micro benchmark under a dedicated system and under anonymous page pressure, respectively. We use the same settings as those in Section~\ref{sec:micro-hdd} to generate the memory pressure. Figures~\ref{fig:small-sensitivity} and \ref{fig:large-sensitivity} show the percentage of latency reduction at specific percentiles for small and large requests, respectively.

Under a dedicated system, a small value of $RSV\_FACTOR$ significantly increases the $99^{th}$ percentile tail latency for small requests, as shown in Figure~\ref{fig:small-sensitivity}(a). The reason is that the reserved memory under such a $RSV\_FACTOR$ value is too small. When a burst of memory requests are sent by the processes, the reserved memory quickly runs out. In this case, the incoming memory requests are blocked by the memory reservation routine.
As the value of $RSV\_FACTOR$ is increased, the $99^{th}$ percentile tail latency becomes better than that by the default Glibc. For large memory requests, the incoming memory requests are not blocked but served by the default allocation routine in Glibc since there can be multiple mmapped memory chunks for a process. Thus, Hermes achieves more allocation latency reduction for large requests under a dedicated system as shown in Figure~\ref{fig:large-sensitivity}(a).

Under anonymous page pressure, Hermes achieves much more significant latency reduction compared with that under a dedicated system. Specifically, it reduces the average and the $99^{th}$ percentile latency by up to 69.1\% and 41.6\% for small requests, respectively. It reduces the average and the $99^{th}$ percentile latency by up to 63.8\% and 64.2\% for large requests, respectively. Overall, setting $RSV\_FACTOR$ to a value larger than 2 does not achieve more performance gain. The reserved memory exceeds the total amount of memory requests and causes more memory wastage. Thus, we empirically set the value of $RSV\_FACTOR$ to 2 since it achieves good reduction in the memory allocation latency while resulting in the least memory wastage.

\subsection{Hermes Overhead}
\label{sec:overhead}


Hermes introduces about 0.4\% CPU usage overhead to the application  due to the management thread in the modified Glibc. We profile the memory that is reserved but not actually used by the micro benchmark for both small (1KB-size) and large (256KB-size) memory requests. 
The reserved memory at runtime is about 6 MB$\sim$6 .4 MB, which is negligible compared to the memory capacity of a physical node. In addition, the memory monitor daemon requires about 2 MB memory including the memory occupied by the daemon process and the shared memory space. It introduces about 2.4\% CPU usage as it keeps monitoring the latency-critical services and available memory in Linux OS.

\section{Discussions}
\label{sec:discussions}

\textbf{Reservation factor.} Users need to set an appropriate value for the reservation factor $RSV\_FACTOR$ in Hermes. We find that a value of 2 achieves good performance gain for both the micro benchmark and two real-world services while introducing the least memory wastage. However, the value setting depends on various factors such as the characteristics of latency-critical services and the underlying multi-tenant system. If a latency-critical service does not require much memory at runtime, $RSV\_FACTOR$ can be set to a small value. Otherwise, it should be set to a relatively large value. 



\noindent
\textbf{Query latency.} Hermes aims for fast memory allocation. Once the reserved pages are obtained by a process, Hermes calls \texttt{munlock} system call on the pages. The pages can be swapped into disks when the available system memory is low. Queries to the latency-critical services will be delayed if the physical pages reside in the swap area. A simple solution is to return the pages to a process without calling \texttt{munlock}. In this case, the pages occupied by latency-critical services are never swapped, resulting in low latency for queries. The simple solution meets the design goal that batch jobs should not affect the performance of latency-critical services. However, it may incur out-of-memory errors if the locked memory is not well managed under extreme memory pressure. Thus, no page is eligible for reclaim but killing processes is the only choice.

\noindent
\textbf{Default Linux mechanisms.} Linux provides mechanisms by which applications can instruct the system to construct the virtual-to-physical memory mapping. For example, the \texttt{MAP\_POPULATE} flag in the \texttt{mmap()} system call and the \texttt{mlock()} system call. However, using these mechanisms in applications has two drawbacks. First, using these mechanisms requires modification to application source code. By doing so, furthermore, an applications bypasses the memory management routine in the library and need to manage the memory by themselves, which put much more burden on software developers. Second, when an application tries to allocate memory under memory pressure, using these mechanisms does not help fast  memory allocation since Linux OS still needs to reclaims/swaps physical memory before the new allocation request.

\textbf{Cgroups.} Linux provides the cgroup mechanism to control resource utilization of processes. Its memory subsystem can be used to dynamically set memory limits for batch jobs. However, there are two limitations by using the cgroup mechanism. First, after the memory of batch jobs is reclaimed by setting a smaller limit in cgroup, the reclaimed memory can be allocated to multiple latency-critical services. This may lead to memory competition and degrades latency. Second, the cgroup mechanism cannot proactively construct the virtual-to-physical mapping.

\noindent
\textbf{Fragmentation.} The current Glibc does not round up the size of heap memory chunks to power of two. Thus, freed memory chunks of any size can be coalesced to neighboring chunks, which does not incur high memory waste through fragmentation. Hermes inherits the heap management algorithm from Glibc for small memory requests allocated from heap. Thus, the impact of fragmentation on heap memory is the same as that in Glibc. Hermes uses its own segregated free list to manage large memory chunks allocated by \texttt{mmap} system call. Since most memory requests from latency-critical services are of the same size, freed large memory chunks may exactly fit incoming requests, incurring no fragmentation. In the worst case where significant memory waste through fragmentation occurs, memory compaction can be done through \texttt{mremap} system call. This is a rare case since modern CPUs support hundreds of gigabytes of memory address space.


\noindent
\textbf{Applicability.} Currently, Hermes supports C/C++ programs. Many popular key-value stores~\cite{Redis, Rocksdb, Memcached, Mongodb} are implemented in C/C++. The principle and design of Hermes can be applied to other language runtimes. For example, for programs running on Java Virtual Machines (JVMs), JVMs could reserve a chunk of memory and construct the virtual-physical mapping in advance for fast memory allocation. 

\section{Related Work}
\label{sec:related-work}

\textbf{Latency reduction.} There are extensive efforts on reducing query latency for latency-critical services~\cite{Hao-SOSP17-MittOS, Hahn-ATC18-FastTrack, Misra-EuroSys19-ManagingTail, Iorgulescu-ATC18-PerfIso, Berger-OSDI18-RobinHood, Li-PPoPP16-tailcontrol, Zhu-SOCC17-WorkloadCompactor, Kogias-Eurosys20-HovercRaft, Gilad-Eurosys20-EvenDB, Chen-Asplos20-FlatStore, Chen-FAST20-HotRing}.
Tail-control~\cite{Li-PPoPP16-tailcontrol} develops a work-stealing scheduler for optimizing the number of requests that meet a target latency. Recently, MittOS~\cite{Hao-SOSP17-MittOS} tackles the tail latency for distributed file systems where the bottleneck is disk I/O. FastTrack~\cite{Hahn-ATC18-FastTrack} targets on mobile devices and improves the response time for foreground apps. PerfIso~\cite{Iorgulescu-ATC18-PerfIso} is an approach that reserves CPU slacks to achieve efficient CPU sharing between latency-critical services and batch jobs. It is orthogonal to Hermes. RobinHood~\cite{Berger-OSDI18-RobinHood} dynamically reallocates the cache between cache-rich and cache-poor applications. CurtailHDFS~\cite{Misra-EuroSys19-ManagingTail} manages the tail latency in distributed file systems. Hermes aims to reduce the latency in the memory allocation phase for latency-critical services in multi-tenant systems and achieve significantly lower tail query latency and higher throughput.

\noindent
\textbf{Cluster resource sharing.} Modern cluster schedulers~\cite{karanasos2015mercury, boutin2014apollo, Chen-ATC17-BIGC} launch best-effort jobs with transient resources in a cluster. For example, Apollo~\cite{boutin2014apollo} is a scalable scheduling framework for cloud computing. Mercury~\cite{karanasos2015mercury} launches jobs with transient resources and kills the jobs when the available resources drop below a threshold. Pado~\cite{Yang-EuroSys17-Pado} is a data processing engine that aims to harness transient resources. Mos~\cite{Anwar-HPDC16-Mos} analyzes cloud object stores and proposes independent microstores for the needs of particular types of workloads. Big-C~\cite{Chen-ATC17-BIGC} is a preemption-based cluster scheduler that achieves low scheduling latency for heterogeneous workloads. 
Hermes targets on efficient co-location and specifically efficient memory sharing in physical nodes.

\noindent
\textbf{Memory management.} Study~\cite{Narasayya-VLDB15-BufferPool} designs a buffer pool for relational databases in a multi-tenant environment. 
There are studies on efficient memory management for applications running in JVMs by leveraging the runtime characteristics of applications~\cite{ROLP-EuroSys19-Bruno, mguyen2015facade, nguyen2016yak, gog2015broom}. 
For example, ROLP~\cite{ROLP-EuroSys19-Bruno} is an object lifetime profiler for efficient garbage collection. Hermes focuses on latency-critical services that use C libraries. 

\noindent
\textbf{Memory allocation libraries.} GNU C Library provides ptmalloc~\cite{Glibc} as the memory allocator for C/C++ programs. There are other memory allocators~\cite{Evans-BSDCan06-jemalloc, Berger-ASPLOS00-Hoard, tcmalloc} that focus on different design objectives. Jemalloc~\cite{Evans-BSDCan06-jemalloc} emphasizes fragmentation avoidance. It is the default memory allocator for FreeBSD~\cite{FreeBSD}. Hoard~\cite{Berger-ASPLOS00-Hoard} is a scalable memory allocator that largely avoids false sharing and is memory efficient. 
TCMalloc~\cite{tcmalloc} supports efficient memory allocation for multi-thread processes. 
Although Hermes is implemented in Glibc, its principle can be integrated to those memory allocators.

\section{Conclusion}
\label{sec:conclusion}
We present Hermes, a library-level mechanism that enables fast memory allocation for latency-critical services in multi-tenant servers. 
Hermes constructs the virtual-physical address mapping in advance and quickly serves incoming memory requests from latency-critical services. It proactively advises Linux OS to release file cache occupied by batch jobs so as to make available memory without going through the slow memory reclaim routine. Hermes is implemented in GNU C Library. 
Experimental results with a micro benchmark and two real-world services show that Hermes significantly reduces the average and the tail latency of queries for latency-critical services especially under memory pressure, and improves system throughput and memory utilization.

In the future, we plan to extend the principle and design of Hermes to language runtimes Java and Scala.

\bibliographystyle{plain}
\bibliography{eddie-simplified}
\end{document}